\documentclass[10pt,twocolumn,tighten,trackchanges]{aastex62}
\usepackage{amsmath}
\usepackage{graphicx}
\usepackage[caption=false]{subfig}
\usepackage{placeins}
\usepackage{float}
\usepackage{comment}

\newcommand{\npsrs}{45}
\newcommand{\gm}{\textrm{gm}}
\newcommand{\Msun}{\ensuremath{\mathrm{M_{\Sun}}}}
\newcommand{\nm}{\textrm{nm}}
\newcommand{\cm}{\textrm{cm}}
\newcommand{\km}{\textrm{km}}
\newcommand{\kpc}{\textrm{kpc}}
\newcommand{\Hz}{\textrm{Hz}}
\newcommand{\MHz}{\textrm{MHz}}
\newcommand{\GHz}{\textrm{GHz}}
\newcommand{\us}{\ensuremath\mathrm{\mu s}}
\newcommand{\ms}{\textrm{ms}}
\newcommand{\s}{\textrm{s}}
\newcommand{\hr}{\textrm{hr}}
\newcommand{\yr}{\textrm{yr}}
\newcommand{\MeV}{\textrm{keV}}
\newcommand{\erg}{\textrm{erg}}
\newcommand{\gauss}{\textrm{G}}
\newcommand{\DM}{\textrm{DM}}
\newcommand{\dmu}{\ensuremath{\mathrm{pc\; cm^{-3}}}}

\begin{document}

\singlespace

\title{The Green Bank North Celestial Cap Pulsar Survey III:
  \npsrs\ New Pulsar Timing Solutions}

\author{Ryan S.\ Lynch}
\affiliation{Green Bank Observatory, PO Box 2, Green Bank, WV
  24494, USA}
\affiliation{Center for Gravitational Waves and Cosmology, Department
  of Physics and Astronomy, West Virginia University, White Hall, Box
  6315, Morgantown, WV 26506, USA}
\author{Joseph K.\ Swiggum}
\affiliation{Center for Gravitation, Cosmology, and Astrophysics,
  Department of Physics, University of Wisconsin-Milwaukee, PO Box
  413, Milwaukee, WI, 53201, USA}
\author{Vlad I.\ Kondratiev}
\affiliation{ASTRON, the Netherlands Institute for Radio Astronomy,
  Postbus 2, 7990 AA, Dwingeloo, The Netherlands}
\affiliation{Astro Space Center of the Lebedev Physical Institute,
  Profsoyuznaya str. 84/32, Moscow, 117997, Russia}
\author{David L.\ Kaplan}
\affiliation{Center for Gravitation, Cosmology, and Astrophysics,
  Department of Physics, University of Wisconsin-Milwaukee, PO Box
  413, Milwaukee, WI, 53201, USA}
\author{Kevin Stovall}
\affiliation{National Radio Astronomy Observatory, 1003 Lopezville
  Rd., Socorro, NM, 87801, USA}
\author{Emmanuel Fonseca}
\affiliation{Department of Physics \& McGill Space Institute, McGill
  University, 3600 University Street, Montreal, QC, H3A 2T8, Canada}
\author{Mallory S.\ E.\ Roberts}
\affiliation{New York University Abu Dhabi, Abu Dhabi, UAE}
\affiliation{Eureka Scientific, Inc., 2452 Delmer St., Suite 100,
  Okland, CA, 94602, USA}
\author{Lina Levin}
\affiliation{Jodrell Bank Centre for Astrophysics, School of Physics
  and Astronomy, The University of Manchester, Manchester, M13 9PL,
  UK}
\author{Megan E.\ DeCesar}
\affiliation{Department of Physics, 730 High St., Lafayette College,
  Easton, PA, 18042, USA}
\author{Bingyi Cui}
\affiliation{Center for Gravitational Waves and Cosmology, Department
  of Physics and Astronomy, West Virginia University, White Hall, Box
  6315, Morgantown, WV 26506, USA}
\author{S.\ Bradley Cenko}
\affiliation{Astrophysics Science Division, NASA Goddard Space Flight
  Center, Mail Code 661, Greenbelt, MD 20771, USA}
\affiliation{Joint Space-Science Institute, University of Maryland,
  College Park, MD 20742, USA}
\author{Pradip Gatkine}
\affiliation{Department of Astronomy, University of Maryland, College
  Park, MD, 20742, USA}
\author{Anne M.\ Archibald}
\affiliation{ASTRON, the Netherlands Institute for Radio Astronomy,
  Postbus 2, 7990 AA, Dwingeloo, The Netherlands}
\author{Shawn Banaszak}
\affiliation{Center for Gravitation, Cosmology, and Astrophysics,
  Department of Physics, University of Wisconsin-Milwaukee, PO Box
  413, Milwaukee, WI, 53201, USA}
\author{Christopher M.\ Biwer}
\affiliation{Department of Physics, Syracuse University, Syracuse, NY,
  13244, USA}
\author{Jason Boyles}
\affiliation{Department of Physics and Astronomy, Western Kentucky
  University, 1906 College Heights Blvd., Bowling Green, KY,
  42101, USA}
\author{Pragya Chawla}
\affiliation{Department of Physics \& McGill Space Institute, McGill
  University, 3600 University Street, Montreal, QC, H3A 2T8, Canada}
\author{Louis P.\ Dartez}
\affiliation{Center for Advanced Radio Astronomy, University of Texas
  Rio Grande Valley, 1 W.\ University Blvd., Brownsville, TX,
  78520, USA}
\author{David Day}
\affiliation{Center for Advanced Radio Astronomy, University of Texas
  Rio Grande Valley, 1 W.\ University Blvd., Brownsville, TX,
  78520, USA}
\author{Anthony J.\ Ford}
\affiliation{Center for Advanced Radio Astronomy, University of Texas
  Rio Grande Valley, 1 W.\ University Blvd., Brownsville, TX,
  78520, USA}
\author{Joseph Flanigan}
\affiliation{Center for Gravitation, Cosmology, and Astrophysics,
  Department of Physics, University of Wisconsin-Milwaukee, PO Box
  413, Milwaukee, WI, 53201, USA}
\author{Jason W.\ T.\ Hessels}
\affiliation{ASTRON, the Netherlands Institute for Radio Astronomy,
  Postbus 2, 7990 AA, Dwingeloo, The Netherlands}
\affiliation{Anton Pannekoek Institute for Astronomy, University of
  Amsterdam, Science Park 904, 1098 XH, Amsterdam, The Netherlands}
\author{Jesus Hinojosa}
\affiliation{Center for Advanced Radio Astronomy, University of Texas
  Rio Grande Valley, 1 W.\ University Blvd., Brownsville, TX,
  78520, USA}
\author{Fredrick A.\ Jenet}
\affiliation{Center for Advanced Radio Astronomy, University of Texas
  Rio Grande Valley, 1 W.\ University Blvd., Brownsville, TX,
  78520, USA}
\author{Chen Karako-Argaman}
\affiliation{Department of Physics \& McGill Space Institute, McGill
  University, 3600 University Street, Montreal, QC, H3A 2T8, Canada}
\author{Victoria M.\ Kaspi}
\affiliation{Department of Physics \& McGill Space Institute, McGill
  University, 3600 University Street, Montreal, QC, H3A 2T8, Canada}
\author{Sean Leake}
\affiliation{Center for Advanced Radio Astronomy, University of Texas
  Rio Grande Valley, 1 W.\ University Blvd., Brownsville, TX,
  78520, USA}
\author{Grady Lunsford}
\affiliation{Center for Advanced Radio Astronomy, University of Texas
  Rio Grande Valley, 1 W.\ University Blvd., Brownsville, TX,
  78520, USA}
\author{Jos\'{e} G.\ Martinez}
\affiliation{Max-Planck-Institut f\"{u} Radioastronomie, Auf dem
  H\"{u}gel 69, D-53121, Bonn, Germany}
\author{Alberto Mata}
\affiliation{Center for Advanced Radio Astronomy, University of Texas
  Rio Grande Valley, 1 W.\ University Blvd., Brownsville, TX,
  78520, USA}
\author{Maura A.\ McLaughlin}
\affiliation{Center for Gravitational Waves and Cosmology, Department
  of Physics and Astronomy, West Virginia University, White Hall, Box
  6315, Morgantown, WV 26506, USA}
\author{Hind Al Noori}
\affiliation{New York University Abu Dhabi, Abu Dhabi, UAE}
\author{Scott M.\ Ransom}
\affiliation{National Radio Astronomy Observatory, 520 Edgemont Road,
Charlottesville, VA, 23903, USA}
\author{Matthew D.\ Rohr}
\affiliation{Center for Gravitation, Cosmology, and Astrophysics,
  Department of Physics, University of Wisconsin-Milwaukee, PO Box
  413, Milwaukee, WI, 53201, USA}
\author{Xavier Siemens}
\affiliation{Center for Gravitation, Cosmology, and Astrophysics,
  Department of Physics, University of Wisconsin-Milwaukee, PO Box
  413, Milwaukee, WI, 53201, USA}
\author{Ren\'ee Spiewak}
\affiliation{Centre for Astrophysics and Supercomputing, Siwnburne
  University of Technology, PO Box 218, Hawthorn, VIC 3122, Australia}
\affiliation{Center for Gravitation, Cosmology, and Astrophysics,
  Department of Physics, University of Wisconsin-Milwaukee, PO Box
  413, Milwaukee, WI, 53201, USA}
\author{Ingrid H.\ Stairs}
\affiliation{Department of Physics and Astronomy, University of
  British Columbia, 6224 Agriculture Rd., Vancouver, BC, V6T 1Z1,
  Canada}
\author{Joeri van Leeuwen}
\affiliation{ASTRON, the Netherlands Institute for Radio Astronomy,
  Postbus 2, 7990 AA, Dwingeloo, The Netherlands}
\affiliation{Anton Pannekoek Institute for Astronomy, University of
  Amsterdam, Science Park 904, 1098 XH, Amsterdam, The Netherlands}
\author{Arielle N.\ Walker}
\affiliation{Center for Gravitation, Cosmology, and Astrophysics,
  Department of Physics, University of Wisconsin-Milwaukee, PO Box
  413, Milwaukee, WI, 53201, USA}
\author{Bradley L.\ Wells}
\affiliation{Department of Atmospheric Sciences, Colorado State
  University, 3915 W.\ Laporte Ave., Fort Collins, CO, 80523, USA}
\affiliation{Center for Gravitation, Cosmology, and Astrophysics,
  Department of Physics, University of Wisconsin-Milwaukee, PO Box
  413, Milwaukee, WI, 53201, USA}

\email{rlynch@nrao.edu}

\shorttitle{GBNCC II: Timing Solutions for \npsrs\ New Pulsars}
\shortauthors{Lynch et al.}

\received{January 7, 2018}
\revised{February 6, 2018}
%\accepted{}
\submitjournal{\apj}

\begin{abstract}
  We provide timing solutions for \npsrs\ radio pulsars discovered by
  the Robert C.\ Byrd Green Bank Telescope.  These pulsars were found
  in the Green Bank North Celestial Cap pulsar survey, an all-GBT-sky
  survey being carried out at a frequency of $350\; \MHz$.  We include
  pulsar timing data from the Green Bank Telescope and Low Frequency
  Array.  Our sample includes five fully recycled millisecond pulsars
  (MSPs, three of which are in a binary system), a new relativistic
  double neutron star system, an intermediate mass binary pulsar, a
  mode-changing pulsar, a 138-ms pulsar with a very low magnetic
  field, and several nulling pulsars.  We have measured two
  post-Keplerian parameters and thus the masses of both objects in the
  double neutron star system.  We also report a tentative companion
  mass measurement via Shapiro delay in a binary MSP.  Two of the MSPs
  can be timed with high precision and have been included in pulsar
  timing arrays being used to search for low-frequency gravitational
  waves, while a third MSP is a member of the black widow class of
  binaries.  Proper motion is measurable in five pulsars and we
  provide an estimate of their space velocity.  We report on an
  optical counterpart to a new black widow system and provide
  constraints on the optical counterparts to other binary MSPs.  We
  also present a preliminary analysis of nulling pulsars in our
  sample.  These results demonstrate the scientific return of long
  timing campaigns on pulsars of all types.
\end{abstract}


\section{Introduction}
\label{sec:intro}

Radio pulsars have long been used as exquisite natural laboratories
for studying a wide range of phenomena in physics and astronomy.  The
well known double neutron star system PSR B1913+16 provided the first
observational evidence for the existence of gravitational waves
\citep[GWs,][]{tw89,wnt10}, and the double pulsar system J0737$-$3039
continues to place ever more stringent constraints on deviations from
general relativity (GR) in the strong-field regime \citep{ksm+06}.
Neutron star mass measurements can be used to study nuclear physics
and the equation-of-state of ultra-dense matter \citep{dpr+10}, while
also providing insight into the mass distribution of the neutron star
population \citep{opns12,kkdt13,ato+16} and in turn formation
mechanisms and evolution \citep[e.g.][]{lp04,ts99}.  Proper motion
measurements can be used to estimate transverse velocity, which also
informs theories of neutron star formation and supernovae energetics
\citep{fsk+13}.  While most binary pulsars have low-mass He white
dwarf (WD) companions, rarer binaries have also been found.  These
include systems with more massive CO WDs \citep[e.g.][]{clm+01},
giant companions that are actively transferring mass to the neutron
star, and so-called black widow binaries where an energetic pulsar
wind has ablated the companion's outer layers, leaving a very low-mass
degenerate core \citep{rob13}.

High-precision millisecond pulsars (MSP, which we define as $P < 10\;
\ms$) are currently being used in an effort to directly detect
nanohertz frequency GWs by forming a pulsar timing array (PTA).  GWs
are predicted to cause nanosecond-scale deviations in pulse arrival
times with a unique angular correlation between pairs of MSPs.  At
nanohertz frequencies the dominant source class is expected to be
supermassive binary black holes in the early stages of inspiral,
though more exotic sources such as cosmic strings are also predicted
to emit at these GW frequencies.  One of the best ways to improve the
sensitivity of PTAs is by adding new high-precision pulsars to the
array, particularly on angular baselines that are not currently well
sampled \citep{sejr13}.  PTAs are a major project at all large radio
telescopes: the North American Nanohertz Observatory for Gravitational
Waves (NANOGrav; \citealt{mcl13}) uses the Robert C.\ Byrd Green Bank
Telescope (GBT) and William E.\ Gordon telescope at the Arecibo
Observatory, the Parkes Pulsar Timing Array (PPTA; \citealt{hob13})
uses the Parkes Telescope, and the European Pulsar Timing Array (EPTA;
\citealt{kd13}) uses the Effelsberg Telescope, the Lovell Telescope,
the Nan\c{c}ay Radio Telescope, the Sardinia Radio Telescope, and the
Westerbork Synthesis Array.  All of these projects also collaborate
under the framework of the International Pulsar Timing Array
\citep{man13}.

Despite many decades of investigation, pulsar emission mechanisms are
not fully understood.  A wide variety of behavior is observed,
however.  This includes abrupt changes in the average pulse profile
between a small number of modes that may be accompanied by changes in
the spin-down rate of the pulsar and are tied to global magnetospheric
reconfigurations \citep{lhk+10}.  Pulsar emission may also be variable
on a variety of timescales: so-called nulling pulsars may cease radio
emission in as short as one rotation, remain in a quiescent state for
many rotations, and then switch back on just as suddenly.  First
discovered by \cite{bac70}, nulling pulsars are an invaluable
population for studying pulsar emission mechanisms and
magnetospheres. Despite nearly 50 years of investigation, nulling
remains poorly understood and only $\sim 130$ pulsars
\citep[$\sim 5\%$ of the known population;][]{gaj17} exhibit nulling
behavior.  In the extreme case of rotating radio transients (RRATs;
\citealt{mll+06}), only a few single pulses are ever observed.

It is essential to conduct long-term timing campaigns on new pulsars
if their scientific impact is to be fully realized.  The arrival times
of a fiducial point in a pulsar's light curve are measured to high
precision and used as input to a model that is coherent in rotational
phase.  Deviations from the predicted arrival times reveal information
about the pulsar and its environment, such as the rotational period
and period derivative, astrometric parameters, the column density of
electrons along the line of sight, and Keplerian and post-Keplerian
orbital parameters.  Because pulsar timing accounts for every rotation
of the pulsar, parameters can be measured with remarkable
precision. The observed rotational parameters can be used to derive
canonical properties such as characteristic age, surface magnetic
field strength, and total spin-down luminosity.  Long-term timing
campaigns can be used to measure even small effects with high
significance, and regular monitoring makes it possible to study
time-variable phenomena such as mode-changing and nulling.

Large-area surveys are the best way to find new and interesting
pulsars.  The Green Bank North Celestial Cap (GBNCC) pulsar survey is
an ongoing $350\; \MHz$ all-sky search for pulsars and transients
being carried out with the GBT \citep{slr+14}.  With 156 pulsars and
RRATs discovered to-date, it is the most successful pulsar survey
conducted in this frequency range.

\begin{deluxetable*}{ccccccc}[t!]
  \centering
  \tabletypesize{\footnotesize}
  \tablewidth{0pt}
  \tablecolumns{7}
  \tablecaption{Observing Set-up \label{table:obs}}
  \tablehead{
    \colhead{Receiver}                              &
    \colhead{$f_{\rm ctr}$ ($\MHz$)} &
    \colhead{$\Delta f$ ($\MHz$)}  &
    \multicolumn{2}{c}{Incoherent Dedispersion}     &
    \multicolumn{2}{c}{Coherent Dedispersion}       \\
    \colhead{}                                      &
    \colhead{}                                      &
    \colhead{}                                      &
    \colhead{$n_{\rm chan}$}         &
    \colhead{$t_{\rm int}$ ($\us$)}  &
    \colhead{$n_{\rm chan}$}                          &
    \colhead{$t_{\rm int}$ ($\us$)}
  }
  \startdata
  LOFAR  & 148  & 78  & 6400 & 327.68 & 400     & 5.12    \\
  PF 342 & 350  & 100 & 4096 & 81.92  & 512     & 10.24   \\
  PF 800 & 820  & 200 & 2048 & 40.96  & 512     & 10.24   \\
  L-Band & 1500 & 800 & 2048 & 40.96  & 512     & 10.24   \\
  S-Band & 2000 & 800 & 2048 & 40.96  & \nodata & \nodata \\
  \enddata
\end{deluxetable*}

In this paper, we report pulsar timing solutions for \npsrs\ pulsars
discovered by the GBNCC survey.  These include five MSPs (two of which
are being timed by NANOGrav and the IPTA), five binary pulsars,
including a relativistic double neutron star (DNS) system, an
intermediate mass binary pulsar, a black widow system, a long-period
pulsar with an anomalously low magnetic field, a mode-changing pulsar,
and six nulling pulsars.  We have measured the masses of both
constituents of the DNS system and also report a tentative mass
measurement via Shapiro delay of a WD companion to an MSP.  An
additional 10 pulsars, including one millisecond pulsar, one disrupted
recycled pulsar, and one nulling pulsar are reported on in
\citet{kmk+18}.  In \S\ref{sec:GBNCC} we provide a brief overview of
the GBNCC survey and in \S\ref{sec:obs} we describe the observational
set-ups that were used in our timing campaign.  In \S\ref{sec:timing}
we provide details of the data reduction and timing analysis.  In
\S\ref{sec:results} we provide timing solutions and discuss select
individual systems, in \S\ref{sec:nullers} we present our nulling
analysis, in \S\ref{sec:optical} we discuss constraints on optical
companions to the binary pulsars, and we summarize our results in
\S\ref{sec:conc}.

\section{The Green Bank North Celestial Cap Pulsar Survey}
\label{sec:GBNCC}

\citet{slr+14} describe the GBNCC survey in detail; here we provide a
brief overview.  The survey began in 2009 covering declinations
$\delta > 38\arcdeg$ and has continued to cover the full sky visible
from the GBT (85\% of the celestial sphere).  It is being carried out
at a center frequency of $350\; \MHz$ and with dwell times of $120\;
\s$.  The low observing frequency preferentially selects sources with
low dispersion measures (DMs, the electron column density) and steep
spectral indices relative to surveys at higher frequencies.  Total
intensity data are collected using the Green Bank Ultimate Pulsar
Processing Instrument (GUPPI) using a bandwidth of $100\; \MHz$, 4096
frequency channels, and $81.92\; \us$ sampling time.  Data are
processed on the Guillimin high performance computer operated by
McGill University, Compute Canada, and Calcul Qu\'{e}bec, using a
pipeline based on the
\texttt{PRESTO}\footnote{\url{http://www.cv.nrao.edu/~sransom/presto}}
software package \citep{rem02}, and are searched for periodic signals
and single pulses.  Candidates are uploaded to a web-based image
viewing and ranking application\footnote{Hosted on
  \url{http://ca.cyberska.org/}}.  Periodicity candidates are analyzed
with a pattern recognition neural net \citep{zbm14}, and single-pulse
candidates are analyzed with a grouping algorithm
\citep{kkl+15,ckj+17}.  Recently, a fast folding algorithm for
periodicity candidates (Parent et al. in prep.) and a neural net
classifier for single-pulse candidates have also been implemented.

To-date, $\sim 75\%$ of the full survey area has been covered.  A
total of 156 pulsars have been discovered\footnote{For an up-to-date
  list, see \url{http://astro.phys.wvu.edu/GBNCC/}}, including 20 MSPs
and 11 RRATs \citep{slr+14,kkl+15}.  Data collection is projected to
finish in 2020.

\section{Observational Set-up and Data Collection}
\label{sec:obs}

We used the GBT and , and the Low Frequency Array (LOFAR) for initial
follow-up and timing.  The observational parameters for each
instrument are presented below and summarized in Table
\ref{table:obs}.

\subsection{GBT Observations}
\label{sec:GBT}

We collected all GBT data with GUPPI in the \texttt{PSRFITS} format.
Dedicated timing observations began in 2013 January, though data from
confirmation observations and test observations during the primary
survey have also been included when available.  All pulsars were
observed regularly for a minimum of one full year, and observations
have continued for select MSPs and binary pulsars.  We used a variety
of observing frequencies and instrumental set-ups, but most data were
collected using the GBT's prime focus receiver at a center frequency
of $820\; \MHz$.  The pulsars' initial positions were only known to a
precision of $\sim 36\arcmin$ (the half-power beamwidth of the GBT at
$350\; \MHz$), so we first obtained a refined position using a
seven-point grid map at $820\; \MHz$ \citep{mhl+02}.  During
preliminary observations, we used GUPPI in its incoherent dedispersion
mode, but as timing solutions improved we also used a coherent
dedispersion mode.  When observing with incoherent dedispersion we
recorded total intensity filterbank data.  In coherent dedispersion
modes we recorded all four Stokes parameters and folded the data in
real-time modulo the instantaneous pulsar period, recording
sub-integrations every 10 seconds.  Observing times varied between
sources and sessions but were typically five to 15 minutes.

\subsection{LOFAR Observations}
\label{sec:LOFAR}

A subset of pulsars discovered in the GBNCC survey, including those
presented in this paper, were also initially followed-up and timed
with LOFAR in the frequency range of 110-188 MHz.  Observations were
carried out for almost two years from 2013 March 6 until 2015 January
14 during LOFAR's Cycles 0-2 (project codes LC0\_022, LC1\_025, and
LC2\_007). For every pulsar one or several gridding observations were
first performed to improve the accuracy of the discovery position to
within 2-3 arcmin by forming coherent tied-array beams (TABs) around
the nominal position \citep[e.g.][]{kkl+15}. In some cases first
timing observations were also carried out with an extra one to two
rings of TABs (7 or 19 beams, ring size of $2.5\arcmin$) to further
refine the position. All observations were conducted with the LOFAR's
Full Core using 42-48 HBA sub-stations in most observations, but not
less than 38 sub-stations. For the timing solutions presented in this
paper only LOFAR timing (not gridding) observations are used. Timing
observations were performed with roughly a monthly cadence. Most
observations were five minutes long, except for the mode changing
pulsar J1628+4406 which was observed for 60 minutes at a time to
increase the chances of catching a transition between the modes.  We
include LOFAR data for 29 of the pulsars presented here.

For the initial gridding and timing observations of slow pulsars and
RRATs we recorded Stokes I data in a $78.125\; \MHz$ band centered at
about $149\; \MHz$ split into 400 subbands (subband numbers
51--450). Each subband in turn was split into 16 channels sampled at
$327.68\; \us$. This setup was also used for timing observation of PSR
J1628+4406.  All subsequent timing observations of slow pulsars and
MSPs (during LC1\_035 and LC2\_007 projects) were carried out
recording 400 subbands of complex-voltage (CV) data sampled at $5.12\;
\us$ in the same frequency range.  LOFAR PULsar Pipeline (PULP) was
run after the observation to dedisperse (coherently for the CV mode)
and clean the data, and fold it using the best available pulsar
ephemeris. The length of sub-integrations was $5\; \s$ for MSPs and
$60\; \s$ for slow pulsars. The folded PSRFITS archive files together
with other pipeline data products including different diagnostic plots
were ingested to the LOFAR's Long Term Archive. For a more detailed
description of the observing setup and PULP see \citet{kvh+16}.

\section{Pulsar Timing Analysis}
\label{sec:timing}

Since the rotational parameters of the pulsars were not initially
known to high precision, we processed early data using \texttt{PRESTO}
to excise radio frequency interference (RFI), search for the pulsars
at each epoch, and fold the data at optimal periods, period
derivatives, and DMs.  When pulsars are in a binary system, orbital
acceleration causes a Doppler modulation of the observed rotational
period that is sinusoidal in the case of nearly circular orbits but
that can take on a more complicated form when the orbital eccentricity
is high.  Binary pulsars were observed at high cadence to sample a
range of orbital phases, and we used a least-squares optimizer to fit
a sinusoid to the observed rotational periods (and period derivatives,
when measurable).  This in turn yielded a low-precision estimate of
the Keplerian orbital parameters, which were used as a starting point
in our timing solutions.  PSR J0509+3801 was soon found to have a high
eccentricity (see \S\ref{sec:J0509}).  Initial orbital parameters were
found by following \citet{bn08} in combination with a least-squares
optimizer that fits non-sinusoidal period modulations.

We calculated pulse times of arrival (TOAs) via Fourier-domain cross
correlation of a noise-free template and the observed pulse profile
\citep{tay92}.  Initial templates were created by fitting one or more
Gaussians to the pulse profiles.  Final templates were made by
phase-coherently adding all high signal-to-noise (S/N) data at a given
frequency and then de-noising the summed profile via wavelet
smoothing.  When there was minimal evolution of the pulse profile
between different observing bands we were able to align templates at
different frequencies.  When this was not possible we allowed for an
arbitrary offset in our timing models between TOAs from different
bands.  At each epoch, we summed the data in time and frequency to
create 60-s sub-integrations and four frequency subbands and
calculated a TOA for each.  In some cases, visual inspection of the
resulting residuals revealed a large number of outliers.  To increase
S/N we further summed the data to obtain single-frequency TOAs, or a
single TOA per observing session.  A small number of individual TOAs
still had very large residuals and were found to suffer from
especially low S/N or, more often, were contaminated by RFI.  These
TOAs were removed from further analysis.

We used the
\texttt{TEMPO}\footnote{\url{http://tempo.sourceforge.net/}} pulsar
timing program to fit a timing model to the observed TOAs.  The basic
model consists of rotational frequency and frequency derivative,
position, and DM.  In some cases we were also able to measure a
significant proper motion.  When appropriate we also fit for Keplerian
and post-Keplerian orbital parameters.  All final timing solutions
were derived using ecliptic coordinates ($\lambda$, $\beta$), which
are nearly orthogonal in timing models and thus provide more accurate
representation of the error ellipse \citep{mnf+16}.  We use the DE430
planetary ephemeris and TT(BIPM) time standard as implemented in
\texttt{TEMPO}.  Most of the binary systems presented here have very
low eccentricity, so we use the ELL1 timing model \citep{lcw+01},
which parameterizes the orbit in terms of the epoch of the ascending
node and the first and second Laplace-Lagrange parameters,
\begin{eqnarray}
  T_{\rm asc} &=& T_0 - P_{\rm b} \frac{\omega}{2 \pi} \\
  \epsilon_1 &=& e \sin{\omega} \\
  \epsilon_2 &=& e \cos{\omega} , 
\end{eqnarray}
where $T_0$ is the epoch of periastron passage, $P_{\rm b}$ is the
binary period, $\omega$ is the longitude of periastron passage, and
$e$ is the eccentricity.  This parameterization is appropriate for
systems where $e^2 a/c \sin{i}$ is much less than errors in TOA
measurements (here, $a$ is the semi-major axis, $i$ is the inclination
angle, and $c$ is the speed of light).  The exception is PSR
J0509+3801, a highly eccentric double neutron star system.

\section{Results}
\label{sec:results}

\begin{figure}[t!]
  \centering
  \includegraphics[width=\columnwidth]{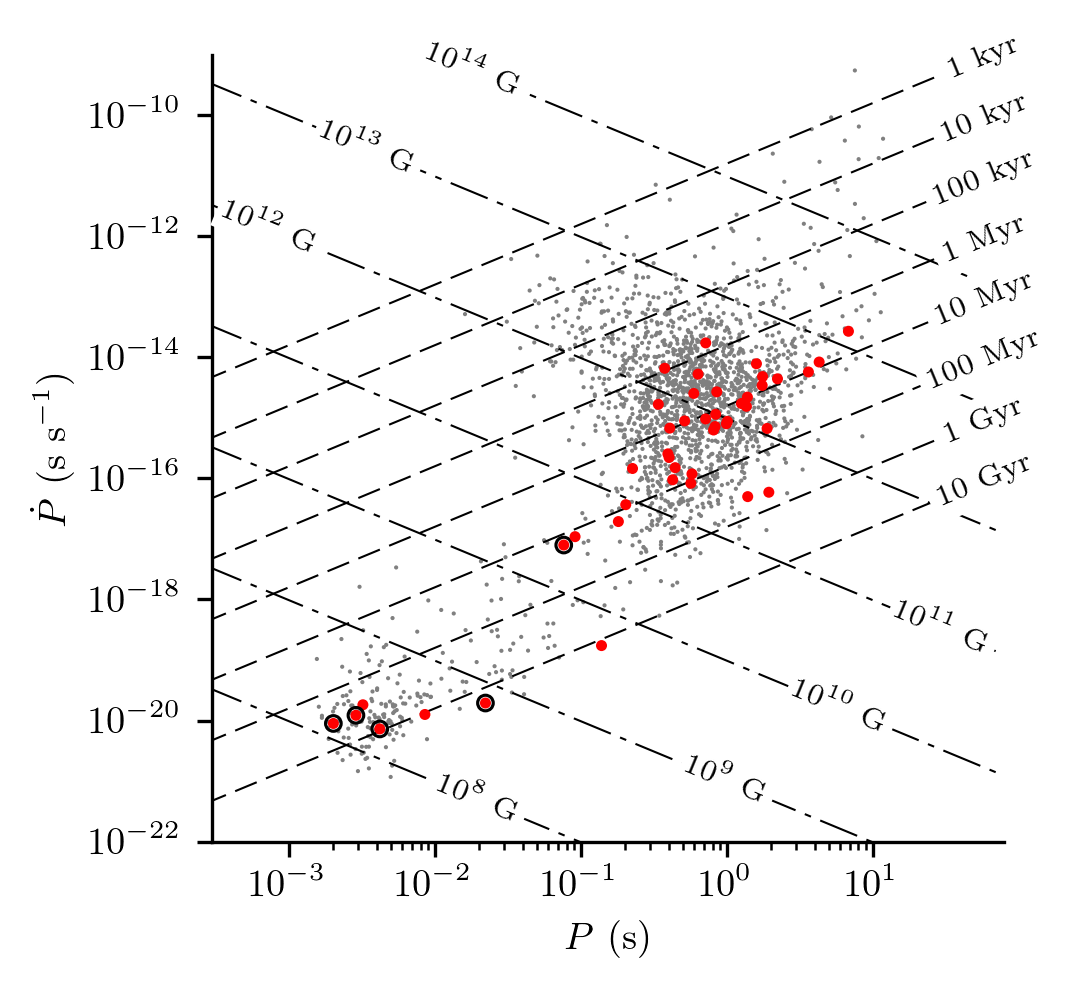}
  \caption{Spin period vs period derivative for the \npsrs\ pulsars
    presented here (red circles) as well as all the pulsars listed in
    the ATNF pulsar catalog\textsuperscript{a} (gray points;
    \citealt{mhth05}).  Binary pulsars are indicated with a black
    circle.  Dashed lines show constant characteristic age and
    dot-dashed lines show constant surface magnetic
    field.  \label{fig:p-pdot}
    \\ \\ \small{\textsuperscript{a}\url{http://www.atnf.csiro.au/research/pulsar/psrcat}}}
\end{figure}

The measured rotational parameters and common derived properties of
all of our pulsars are shown in Tables \ref{table:rotational_measured}
and \ref{table:common_derived}, respectively.  Table
\ref{table:astrometric_dm} shows positions in a variety of coordinate
systems as well as DMs.  ELL1 binary timing parameters of four pulsars
are shown in Table \ref{table:binary}.  The locations of
all the pulsars presented here on the $P$-$\dot{P}$ diagram are shown
in Figure \ref{fig:p-pdot}.  Post-fit timing residuals and integrated
pulse profiles are shown in Figures \ref{fig:residuals} and
\ref{fig:profiles}, respectively. In the following sections, we
present specific details on pulsars of interest.

\subsection{NANOGrav and IPTA Pulsars}
\label{sec:ptas}

One of the primary science goals of the GBNCC survey is finding
high-precision MSPs for inclusion in PTAs.  The quadrupolar nature of
GWs should cause a unique angular correlation between pairs of MSPs in
a PTA.  Sampling many baselines is essential to firmly establish this
quadrupolar signature, and this will be necessary for PTAs to claim a
detection of low-frequency GWs.  The GBNCC survey is especially well
suited to find MSPs at high northern declinations, which have been
historically undersampled by sensitive pulsar surveys.

Two MSPs presented here have been included in NANOGrav and the IPTA
based on promising early results: PSRs J0740+6620 and J1125+7819.  We
found it necessary to use the DMX timing model for PSR J1125+7819 to
measure epoch-dependent variations in DM.  We used a 30-day window for
DMX epochs and found typical DM variations of $\sim
10^{-4}$--$10^{-5}\; \dmu$, with a maximum $| \Delta \DM | = 1.7
\times 10^{-2}\; \dmu$.  These results are consistent with DM
variations measured in NANOGrav timing\footnote{NANOGrav observes PSR
  J1125+7819 at $820$ and $1500\; \MHz$ using coherent dedispersion,
  with 30 minute integrations per TOA.  Nearly all of our timing data
  was taken at $350$ and $820\; \MHz$ but with incoherent dedispersion
  and shorter integration times.  As such, NANOGrav achieves better
  RMS timing residuals.} \citep{abb+18}.  PSRs J1710+4923 and
J1641+8049 have not been included in PTAs despite relatively low
uncertainty on individual TOAs---PSR J1710+4923 because of strong
scintillation and PSR J1641+8049 because it is in a black widow binary
system.

\begin{deluxetable*}{lcrrccccc}[t!]
  \tabletypesize{\footnotesize}
  \tablewidth{\textwidth}
  \tablecaption{Rotational and Timing Parameters of GBNCC
    Pulsars \label{table:rotational_measured}}
  \tablecolumns{9}
  \tablehead{
    \colhead{PSR} & \colhead{Name Used In} & \colhead{$\nu$} & 
    \colhead{$\dot{\nu}$} &
    \colhead{Epoch} & \colhead{Data Span} & 
    \colhead{RMS Residual} & \colhead{$N_{\rm TOA}$} & 
    \colhead{$\chi^2_{\rm red}$} \\ 
    \colhead{} & \colhead{\citet{slr+14}} & \colhead{($\Hz$)} & 
    \colhead{($\Hz\; \s^{-1}$)} & 
    \colhead{(MJD)} & \colhead{(MJD)} & \colhead{($\us$)} & \colhead{}}
  \startdata
  J0054$+$6946 & J0053+69 & 1.200607994501(6) & $-$1.037(1)$\times$10$^{-15}$ & 56500.0 & 56329--56791 & 542.59 & 297 & 0.9 \\
  J0058$+$4950 & J0059+50 & 1.00399242379(4) & $-$8.1(2)$\times$10$^{-16}$ & 56500.0 & 56358--56670 & 1701.9 & 186 & 1.9 \\
  J0111$+$6624 & J0112+66 & 0.23245693423(2) & $-$4.52(9)$\times$10$^{-16}$ & 56500.0 & 56332--56670 & 3071.21 & 131 & 1.6 \\
  J0137$+$6349 & J0136+63 & 1.39282996541(1) & $-$1.862(9)$\times$10$^{-15}$ & 56500.0 & 56329--56670 & 797.51 & 276 & 1.3 \\
  J0212$+$5222 & J0213+52 & 2.65684439046(1) & $-$4.6583(9)$\times$10$^{-14}$ & 56500.0 & 56332--56670 & 376.92 & 252 & 1.6 \\
  J0325$+$6744 & J0325+67 & 0.732773180015(8) & $-$8.34(5)$\times$10$^{-16}$ & 56500.0 & 56351--56669 & 952.49 & 204 & 2.6 \\
  J0335$+$6623 & J0338+66 & 0.56755806947(6) & $-$1.55(3)$\times$10$^{-15}$ & 56500.0 & 56351--56669 & 1068.56 & 88 & 1.2 \\
  J0358$+$4155 & J0358+42 & 4.415316459453(4) & $-$2.843(3)$\times$10$^{-15}$ & 56500.0 & 56332--56670 & 85.32 & 247 & 1.1 \\
  J0358$+$6627 & J0358+66 & 10.92827132616(2) & $-$1.30(1)$\times$10$^{-15}$ & 56500.0 & 56329--56669 & 80.73 & 86 & 0.9 \\
  J0509$+$3801 & J0510+38 & 13.06483380156(2) & $-$1.3538(4)$\times$10$^{-15}$ & 56900.0 & 56336--57474 & 102.36 & 791 & 1.1 \\
  J0518$+$5416 & J0519+54 & 2.93942447332(5) & $-$1.448(3)$\times$10$^{-14}$ & 56500.0 & 56336--56669 & 961.97 & 207 & 3.4 \\
  J0612$+$3721 & J0610+37 & 2.2529049469(1) & $-$7.6(7)$\times$10$^{-16}$ & 56500.0 & 56336--56669 & 1373.35 & 69 & 1.0 \\
  J0738$+$6904 & J0737+69 & 0.14646235983(1) & $-$5.785(8)$\times$10$^{-16}$ & 56425.0 & 56159--56669 & 2138.93 & 84 & 4.0 \\
  J0740$+$6620 & J0741+66 & 346.53199660394(1) & $-$1.4658(8)$\times$10$^{-15}$ & 56675.0 & 56156--56908 & 2.95 & 383 & 1.9 \\
  J0747$+$6646 & J0746+66 & 2.45278075482(3) & $-$4.08(2)$\times$10$^{-15}$ & 56500.0 & 56336--56669 & 371.01 & 73 & 0.9 \\
  J0944$+$4106 & J0943+41 & 0.448544888919(9) & $-$8.894(6)$\times$10$^{-16}$ & 56350.0 & 56046--56669 & 515.44 & 165 & 2.2 \\
  J1059$+$6459 & J1101+65 & 0.2753933549(1) & $-$4.35(4)$\times$10$^{-16}$ & 56500.0 & 56478--57215 & 4371.22 & 30 & 0.8 \\
  J1125$+$7819 & J1122+78 & 238.00405319610(8) & $-$4.11(5)$\times$10$^{-16}$ & 56625.0 & 56156--57078 & 11.58 & 1461 & 3.8 \\
  J1624$+$8643 & J1627+86 & 2.52676461497(1) & $-$1.625(6)$\times$10$^{-15}$ & 56500.0 & 56351--56669 & 223.5 & 279 & 1.1 \\
  J1628$+$4406 & J1629+43 & 5.519419001073(2) & $-$5.9032(8)$\times$10$^{-16}$ & 56625.0 & 56017--56973 & 99.14 & 2225 & 3.7 \\
  J1641$+$8049 & J1649+80 & 494.76063707093(2) & $-$2.19(2)$\times$10$^{-15}$ & 56425.0 & 56159--56670 & 5.81 & 788 & 2.7 \\
  J1647$+$6608 & J1647+66 & 0.62507876955(1) & $-$3.063(6)$\times$10$^{-15}$ & 56500.0 & 56351--56669 & 690.08 & 174 & 1.2 \\
  J1710$+$4923 & J1710+49 & 310.536979440897(4) & $-$1.7561(2)$\times$10$^{-15}$ & 56850.0 & 55996--57700 & 10.76 & 757 & 2.3 \\
  J1800$+$5034 & J1800+50 & 1.72898175945(3) & $-$3.5(2)$\times$10$^{-16}$ & 56500.0 & 56365--56670 & 814.31 & 338 & 1.2 \\
  J1815$+$5546 & J1815+55 & 2.3427772328(2) & $-$5.2(8)$\times$10$^{-16}$ & 56500.0 & 56365--56670 & 1800.71 & 103 & 4.5 \\
  J1821$+$4147 & J1821+41 & 0.792482693574(5) & $-$1.0860(6)$\times$10$^{-15}$ & 56375.0 & 56085--56670 & 625.93 & 252 & 1.1 \\
  J1859$+$7654 & J1859+76 & 0.71749953039(1) & $-$2.6(8)$\times$10$^{-17}$ & 56500.0 & 56351--56669 & 682.37 & 185 & 1.0 \\
  J1923$+$4243 & J1921+42 & 1.68012768584(3) & $-$7.15(1)$\times$10$^{-15}$ & 56525.0 & 56365--56670 & 520.97 & 228 & 1.1 \\
  J1934$+$5219 & J1935+52 & 1.75919355395(8) & $-$2.6(4)$\times$10$^{-16}$ & 56525.0 & 56365--56670 & 2086.41 & 178 & 2.3 \\
  J1938$+$6604 & J1939+66 & 44.926049244297(5) & $-$3.93(7)$\times$10$^{-17}$ & 56550.0 & 56176--56907 & 28.02 & 691 & 1.4 \\
  J1941$+$4320 & J1941+43 & 1.189192897888(8) & $-$1.6224(7)$\times$10$^{-15}$ & 56375.0 & 56081--56670 & 848.81 & 286 & 1.9 \\
  J1942$+$8106 & J1942+81 & 4.91259383196(3) & $-$8.9(1)$\times$10$^{-16}$ & 56500.0 & 56332--56670 & 274.58 & 115 & 1.3 \\
  J1954$+$4357 & J1954+43 & 0.72095913858(1) & $-$1.145(2)$\times$10$^{-15}$ & 56550.0 & 56081--57015 & 1485.51 & 440 & 1.5 \\
  J1955$+$6708 & J1953+67 & 116.74869454326(1) & $-$1.717(6)$\times$10$^{-16}$ & 56600.0 & 56156--57447 & 48.87 & 564 & 3.0 \\
  J2001$+$4258 & J2001+42 & 1.390499284334(8) & $-$3.3501(4)$\times$10$^{-14}$ & 56500.0 & 56328--56670 & 410.55 & 250 & 1.1 \\
  J2017$+$5906 & J2017+59 & 2.4784478211(1) & $-$1.361(8)$\times$10$^{-15}$ & 56575.0 & 56276--56856 & 3397.28 & 321 & 2.4 \\
  J2027$+$7502 & J2027+74 & 1.94092353149(3) & $-$3.35(2)$\times$10$^{-15}$ & 56500.0 & 56351--56669 & 1371.4 & 335 & 1.5 \\
  J2123$+$5434 & J2122+54 & 7.20107927874(1) & $-$9.0(5)$\times$10$^{-18}$ & 56500.0 & 56332--57815 & 92.15 & 249 & 1.0 \\
  J2137$+$6428 & J2137+64 & 0.57110576181(2) & $-$1.12(1)$\times$10$^{-15}$ & 56500.0 & 56329--56670 & 1182.98 & 65 & 1.1 \\
  J2208$+$4056 & J2207+40 & 1.56996372132(3) & $-$1.3022(2)$\times$10$^{-14}$ & 56375.0 & 56081--56670 & 1882.71 & 235 & 1.9 \\
  J2228$+$6447 & J2229+64 & 0.52826709593(1) & $-$1.87(8)$\times$10$^{-16}$ & 56500.0 & 56329--56670 & 1868.72 & 209 & 1.1 \\
  J2241$+$6941 & J2243+69 & 1.16904199244(2) & $-$3.68(1)$\times$10$^{-15}$ & 56500.0 & 56351--56670 & 889.58 & 173 & 1.4 \\
  J2310$+$6706 & \nodata & 0.5141946062(3) & $-$2(1)$\times$10$^{-17}$ & 57225.0 & 57078--57397 & 1204.93 & 80 & 2.5 \\
  J2312$+$6931 & J2316+69 & 1.22944554793(3) & $-$9.5(2)$\times$10$^{-16}$ & 56500.0 & 56331--56670 & 1657.42 & 146 & 2.4 \\
  J2351$+$8533 & J2353+85 & 0.98840874179(6) & $-$8.6(3)$\times$10$^{-16}$ & 56500.0 & 56332--56669 & 1398.72 & 42 & 1.1 \\
  \enddata
  \tablecomments{Where aplicable, we give the pulsar name used when
  presenting discovery parameters in \citet{slr+14}.  All timing
  models use the DE430 Solar system ephemeris and are referenced to
  the TT(BIPM) time standard.  Values in parentheses are the
  $1$-$\sigma$ uncertainty in the last digit as reported by
  \texttt{TEMPO}.}
\end{deluxetable*}

\begin{deluxetable*}{lrrrrrcc}[h!]
  \tabletypesize{\footnotesize}
  \tablewidth{\textwidth}
  \tablecaption{Derived Common Properties of GBNCC Pulsars
    \label{table:common_derived}}
  \tablecolumns{8}
  \tablehead{
    \colhead{PSR} & \colhead{$P$} & \colhead{$\dot{P}$} &
    \colhead{$\tau_{\rm c}$} & \colhead{$B_{\rm surf}$} &
    \colhead{$\dot{E}_{\rm rot}$} & \colhead{$D^{\rm NE2001}_{\rm DM}$} &
    \colhead{$D^{\rm YMW16}_{\rm DM}$} \\
    \colhead{} & \colhead{($\s$)} & \colhead{($\s\; \s^{-1}$)} &
    \colhead{($\yr$)} &  \colhead{($\gauss$)} & 
    \colhead{($\erg\; \s^{-1}$)} & \colhead{($\kpc$)} & \colhead{($\kpc$)}}
    \startdata
  J0054$+$6946 & 0.832911328744(4) & 7.194(8)$\times$10$^{-16}$ & 1.834(2)$\times$10$^{7}$ & 7.832(4)$\times$10$^{11}$ & 4.915(5)$\times$10$^{31}$ & 4.3 & 2.8 \\
  J0058$+$4950 & 0.99602345227(4) & 8.0(2)$\times$10$^{-16}$ & 1.97(5)$\times$10$^{7}$ & 9.0(1)$\times$10$^{11}$ & 3.20(7)$\times$10$^{31}$ & 2.8 & 2.6 \\
  J0111$+$6624 & 4.3018721007(3) & 8.4(2)$\times$10$^{-15}$ & 8.1(2)$\times$10$^{6}$ & 6.07(6)$\times$10$^{12}$ & 4.15(9)$\times$10$^{30}$ & 3.4 & 2.4 \\
  J0137$+$6349 & 0.717962726847(7) & 9.60(5)$\times$10$^{-16}$ & 1.185(6)$\times$10$^{7}$ & 8.40(2)$\times$10$^{11}$ & 1.024(5)$\times$10$^{32}$ & 44.3 & 9.1 \\
  J0212$+$5222 & 0.376386364060(2) & 6.599(1)$\times$10$^{-15}$ & 9.037(2)$\times$10$^{5}$ & 1.5946(1)$\times$10$^{12}$ & 4.8860(9)$\times$10$^{33}$ & 1.5 & 1.6 \\
  J0325$+$6744 & 1.36467876728(1) & 1.553(9)$\times$10$^{-15}$ & 1.393(8)$\times$10$^{7}$ & 1.473(4)$\times$10$^{12}$ & 2.41(1)$\times$10$^{31}$ & 2.3 & 1.9 \\
  J0335$+$6623 & 1.7619342474(2) & 4.81(9)$\times$10$^{-15}$ & 5.8(1)$\times$10$^{6}$ & 2.95(3)$\times$10$^{12}$ & 3.47(7)$\times$10$^{31}$ & 2.3 & 1.9 \\
  J0358$+$4155 & 0.2264843322519(2) & 1.458(1)$\times$10$^{-16}$ & 2.461(2)$\times$10$^{7}$ & 1.8388(9)$\times$10$^{11}$ & 4.956(5)$\times$10$^{32}$ & 1.6 & 1.5 \\
  J0358$+$6627 & 0.0915057807547(2) & 1.09(1)$\times$10$^{-17}$ & 1.33(1)$\times$10$^{8}$ & 3.20(2)$\times$10$^{10}$ & 5.62(5)$\times$10$^{32}$ & 2.2 & 1.9 \\
  J0509$+$3801 & 0.0765413487220(1) & 7.931(2)$\times$10$^{-18}$ & 1.5291(5)$\times$10$^{8}$ & 2.4929(4)$\times$10$^{10}$ & 6.982(2)$\times$10$^{32}$ & 1.9 & 1.6 \\
  J0518$+$5416 & 0.340202651599(6) & 1.676(3)$\times$10$^{-15}$ & 3.217(6)$\times$10$^{6}$ & 7.639(7)$\times$10$^{11}$ & 1.680(3)$\times$10$^{33}$ & 1.5 & 1.4 \\
  J0612$+$3721 & 0.44387136767(2) & 1.5(1)$\times$10$^{-16}$ & 4.7(4)$\times$10$^{7}$ & 2.6(1)$\times$10$^{11}$ & 6.8(6)$\times$10$^{31}$ & 1.2 & 1.1 \\
  J0738$+$6904 & 6.8276928023(5) & 2.697(4)$\times$10$^{-14}$ & 4.012(6)$\times$10$^{6}$ & 1.373(1)$\times$10$^{13}$ & 3.345(5)$\times$10$^{30}$ & 0.8 & 1.1 \\
  J0740$+$6620 & 0.0028857364104907(1) & 1.2206(7)$\times$10$^{-20}$ & 3.746(2)$\times$10$^{9}$ & 1.8990(5)$\times$10$^{8}$ & 2.005(1)$\times$10$^{34}$ & 0.7 & 0.9 \\
  J0747$+$6646 & 0.407700524409(4) & 6.78(3)$\times$10$^{-16}$ & 9.53(4)$\times$10$^{6}$ & 5.32(1)$\times$10$^{11}$ & 3.95(2)$\times$10$^{32}$ & 1.2 & 1.9 \\
  J0944$+$4106 & 2.22943126698(4) & 4.421(3)$\times$10$^{-15}$ & 7.990(6)$\times$10$^{6}$ & 3.176(1)$\times$10$^{12}$ & 1.575(1)$\times$10$^{31}$ & 0.8 & 2.7 \\
  J1059$+$6459 & 3.631169678(2) & 5.73(5)$\times$10$^{-15}$ & 1.003(9)$\times$10$^{7}$ & 4.62(2)$\times$10$^{12}$ & 4.73(4)$\times$10$^{30}$ & 0.8 & 1.9 \\
  J1125$+$7819 & 0.004201609117875(1) & 7.26(8)$\times$10$^{-21}$ & 9.2(1)$\times$10$^{9}$ & 1.77(1)$\times$10$^{8}$ & 3.87(4)$\times$10$^{33}$ & 0.6 & 0.8 \\
  J1624$+$8643 & 0.395763022038(2) & 2.55(1)$\times$10$^{-16}$ & 2.46(1)$\times$10$^{7}$ & 3.211(6)$\times$10$^{11}$ & 1.621(6)$\times$10$^{32}$ & 3.0 & 0.5 \\
  J1628$+$4406 & 0.18117848994714(5) & 1.9378(3)$\times$10$^{-17}$ & 1.4814(2)$\times$10$^{8}$ & 5.9952(4)$\times$10$^{10}$ & 1.2863(2)$\times$10$^{32}$ & 0.6 & 8.8 \\
  J1641$+$8049 & 0.0020211793846822(1) & 8.95(7)$\times$10$^{-21}$ & 3.58(3)$\times$10$^{9}$ & 1.361(5)$\times$10$^{8}$ & 4.28(3)$\times$10$^{34}$ & 1.7 & 2.1 \\
  J1647$+$6608 & 1.59979837535(3) & 7.84(2)$\times$10$^{-15}$ & 3.233(7)$\times$10$^{6}$ & 3.583(4)$\times$10$^{12}$ & 7.56(2)$\times$10$^{31}$ & 1.3 & 3.0 \\
  J1710$+$4923 & 0.00322022839856445(4) & 1.8211(2)$\times$10$^{-20}$ & 2.8017(3)$\times$10$^{9}$ & 2.4502(1)$\times$10$^{8}$ & 2.1530(2)$\times$10$^{34}$ & 0.7 & 0.5 \\
  J1800$+$5034 & 0.57837510115(1) & 1.19(5)$\times$10$^{-16}$ & 7.7(4)$\times$10$^{7}$ & 2.65(6)$\times$10$^{11}$ & 2.4(1)$\times$10$^{31}$ & 1.4 & 1.9 \\
  J1815$+$5546 & 0.42684382706(4) & 9(2)$\times$10$^{-17}$ & 7(1)$\times$10$^{7}$ & 2.0(2)$\times$10$^{11}$ & 4.8(8)$\times$10$^{31}$ & 50.0 & 25.0 \\
  J1821$+$4147 & 1.261857209133(9) & 1.7292(9)$\times$10$^{-15}$ & 1.1562(6)$\times$10$^{7}$ & 1.4946(4)$\times$10$^{12}$ & 3.398(2)$\times$10$^{31}$ & 2.5 & 4.4 \\
  J1859$+$7654 & 1.39372913521(2) & 5(2)$\times$10$^{-17}$ & 4(1)$\times$10$^{8}$ & 2.7(4)$\times$10$^{11}$ & 6(2)$\times$10$^{29}$ & 2.9 & 6.0 \\
  J1923$+$4243 & 0.59519285851(1) & 2.534(5)$\times$10$^{-15}$ & 3.721(8)$\times$10$^{6}$ & 1.243(1)$\times$10$^{12}$ & 4.75(1)$\times$10$^{32}$ & 3.2 & 4.7 \\
  J1934$+$5219 & 0.56844228297(3) & 8(1)$\times$10$^{-17}$ & 1.1(2)$\times$10$^{8}$ & 2.2(2)$\times$10$^{11}$ & 1.8(3)$\times$10$^{31}$ & 4.5 & 7.7 \\
  J1938$+$6604 & 0.022258801226038(2) & 1.95(3)$\times$10$^{-20}$ & 1.81(3)$\times$10$^{10}$ & 6.66(6)$\times$10$^{8}$ & 7.0(1)$\times$10$^{31}$ & 2.3 & 3.4 \\
  J1941$+$4320 & 0.840906468392(6) & 1.1472(5)$\times$10$^{-15}$ & 1.1614(5)$\times$10$^{7}$ & 9.938(2)$\times$10$^{11}$ & 7.617(3)$\times$10$^{31}$ & 4.4 & 6.5 \\
  J1942$+$8106 & 0.203558452868(1) & 3.68(6)$\times$10$^{-17}$ & 8.8(1)$\times$10$^{7}$ & 8.76(7)$\times$10$^{10}$ & 1.72(3)$\times$10$^{32}$ & 2.1 & 3.5 \\
  J1954$+$4357 & 1.38704116015(2) & 2.202(3)$\times$10$^{-15}$ & 9.98(1)$\times$10$^{6}$ & 1.768(1)$\times$10$^{12}$ & 3.258(4)$\times$10$^{31}$ & 7.1 & 5.3 \\
  J1955$+$6708 & 0.0085654062678141(8) & 1.259(4)$\times$10$^{-20}$ & 1.078(4)$\times$10$^{10}$ & 3.323(6)$\times$10$^{8}$ & 7.91(3)$\times$10$^{32}$ & 3.4 & 10.1 \\
  J2001$+$4258 & 0.719166137852(4) & 1.7327(2)$\times$10$^{-14}$ & 6.5762(8)$\times$10$^{5}$ & 3.5717(2)$\times$10$^{12}$ & 1.8390(2)$\times$10$^{33}$ & 3.3 & 3.8 \\
  J2017$+$5906 & 0.40347833490(2) & 2.22(1)$\times$10$^{-16}$ & 2.88(2)$\times$10$^{7}$ & 3.026(9)$\times$10$^{11}$ & 1.332(8)$\times$10$^{32}$ & 3.3 & 3.9 \\
  J2027$+$7502 & 0.515218649152(8) & 8.90(5)$\times$10$^{-16}$ & 9.17(5)$\times$10$^{6}$ & 6.85(2)$\times$10$^{11}$ & 2.57(1)$\times$10$^{32}$ & 1.0 & 0.8 \\
  J2123$+$5434 & 0.1388680725891(2) & 1.7(1)$\times$10$^{-19}$ & 1.27(7)$\times$10$^{10}$ & 5.0(1)$\times$10$^{9}$ & 2.6(1)$\times$10$^{30}$ & 2.1 & 1.8 \\
  J2137$+$6428 & 1.75098916326(7) & 3.43(4)$\times$10$^{-15}$ & 8.1(1)$\times$10$^{6}$ & 2.48(1)$\times$10$^{12}$ & 2.53(3)$\times$10$^{31}$ & 4.8 & 3.8 \\
  J2208$+$4056 & 0.63695739361(1) & 5.283(1)$\times$10$^{-15}$ & 1.9102(4)$\times$10$^{6}$ & 1.8561(2)$\times$10$^{12}$ & 8.071(2)$\times$10$^{32}$ & 1.0 & 0.8 \\
  J2228$+$6447 & 1.89298180355(5) & 6.7(3)$\times$10$^{-16}$ & 4.5(2)$\times$10$^{7}$ & 1.14(3)$\times$10$^{12}$ & 3.9(2)$\times$10$^{30}$ & 46.9 & 6.9 \\
  J2241$+$6941 & 0.85540126571(1) & 2.693(8)$\times$10$^{-15}$ & 5.03(1)$\times$10$^{6}$ & 1.536(2)$\times$10$^{12}$ & 1.699(5)$\times$10$^{32}$ & 2.9 & 2.5 \\
  J2310$+$6706 & 1.944788973(1) & 6(5)$\times$10$^{-17}$ & 5(4)$\times$10$^{8}$ & 3(1)$\times$10$^{11}$ & 2(3)$\times$10$^{29}$ & 3.5 & 2.7 \\
  J2312$+$6931 & 0.81337477832(2) & 6.3(1)$\times$10$^{-16}$ & 2.04(4)$\times$10$^{7}$ & 7.25(6)$\times$10$^{11}$ & 4.63(8)$\times$10$^{31}$ & 2.8 & 2.4 \\
  J2351$+$8533 & 1.01172719111(6) & 8.8(3)$\times$10$^{-16}$ & 1.83(7)$\times$10$^{7}$ & 9.5(2)$\times$10$^{11}$ & 3.3(1)$\times$10$^{31}$ & 1.9 & 2.6 \\
  \enddata
  \tablecomments{$D_{\rm DM}$ is calculated using the NE2001
  \cite{cl02} or YMW16 \citep{ymw17} Galactic free electron density
  models, as indicated.  A fractional uncertainty of 50\% is not
  uncommon.  Derived parameters have not been corrected for the
  Shklovskii effect.  $\dot{E}$ is calculated assuming a moment of
  inertia $I = 10^{45}\; \gm\, \cm^2$. Values in parentheses are the
  $1$-$\sigma$ uncertainty in the last digit, calculated by
  propogating uncertainties in measured parameters reported by
  \texttt{TEMPO}.}
  \end{deluxetable*}

\begin{deluxetable*}{lrrrcccc}[h!]
  \tabletypesize{\footnotesize}
  \tablewidth{\textwidth}
  \tablecaption{Coordinates and DMs of GBNCC Pulsars
   \label{table:astrometric_dm}}
  \tablecolumns{8}
  \tablehead{\colhead{PSR} & \multicolumn{3}{c}{Measured} & 
             \multicolumn{4}{c}{Derived} \\ 
             \colhead{} & \colhead{$\lambda$ ($\arcdeg$)} & 
             \colhead{$\beta$ ($\arcdeg$)} &
             \colhead{DM ($\dmu$)} & 
             \colhead{$\alpha$ (J2000)} & \colhead{$\delta$ (J2000)} & 
             \colhead{$\ell$ ($\arcdeg$)} & \colhead{$b$ ($\arcdeg$)}}
  \startdata
  J0054$+$6946 & 53.17957(2) & 55.915268(8) & 116.52(5) & $00^{\rm h}\, 54^{\rm m}\, 59\, \fs109$ & $69\arcdeg\, 46\arcmin\, 16\, \farcs9$ & 123.24077 & 6.90173 \\
  J0058$+$4950 & 35.9410(1) & 39.55191(6) & 66.953(7) & $00^{\rm h}\, 58^{\rm m}\, 09\, \fs989$ & $49\arcdeg\, 50\arcmin\, 26\, \farcs0$ & 124.04528 & -13.01663 \\
  J0111$+$6624 & 51.3828(4) & 52.3710(1) & 111.20(3) & $01^{\rm h}\, 11^{\rm m}\, 21\, \fs862$ & $66\arcdeg\, 24\arcmin\, 10\, \farcs9$ & 124.92791 & 3.60854 \\
  J0137$+$6349 & 52.48260(5) & 48.69309(3) & 285.50(6) & $01^{\rm h}\, 37^{\rm m}\, 13\, \fs362$ & $63\arcdeg\, 49\arcmin\, 34\, \farcs4$ & 127.95520 & 1.40191 \\
  J0212$+$5222 & 50.60763(2) & 36.42041(1) & 38.21(3) & $02^{\rm h}\, 12^{\rm m}\, 52\, \fs136$ & $52\arcdeg\, 22\arcmin\, 49\, \farcs5$ & 135.33111 & -8.51996 \\
  J0325$+$6744 & 69.66385(5) & 47.02620(3) & 65.28(5) & $03^{\rm h}\, 25^{\rm m}\, 05\, \fs117$ & $67\arcdeg\, 44\arcmin\, 59\, \farcs4$ & 136.71527 & 9.08929 \\
  J0335$+$6623 & 70.4110(3) & 45.3827(3) & 66.726(2) & $03^{\rm h}\, 35^{\rm m}\, 57\, \fs077$ & $66\arcdeg\, 23\arcmin\, 23\, \farcs6$ & 138.38339 & 8.58769 \\
  J0358$+$4155 & 66.154390(4) & 20.973526(6) & 46.325(1) & $03^{\rm h}\, 58^{\rm m}\, 03\, \fs174$ & $41\arcdeg\, 55\arcmin\, 19\, \farcs1$ & 156.11209 & -8.62040 \\
  J0358$+$6627 & 73.496599(8) & 44.753531(6) & 62.33(1) & $03^{\rm h}\, 58^{\rm m}\, 37\, \fs926$ & $66\arcdeg\, 27\arcmin\, 46\, \farcs6$ & 140.12843 & 10.06385 \\
  J0509$+$3801 & 79.7362277(8) & 15.030898(4) & 69.0794(9) & $05^{\rm h}\, 09^{\rm m}\, 31\, \fs788$ & $38\arcdeg\, 01\arcmin\, 18\, \farcs1$ & 168.27474 & -1.18699 \\
  J0518$+$5416 & 83.01300(5) & 31.09029(8) & 42.330(5) & $05^{\rm h}\, 18^{\rm m}\, 53\, \fs198$ & $54\arcdeg\, 16\arcmin\, 50\, \farcs0$ & 155.91644 & 9.55747 \\
  J0612$+$3721 & 92.6070(1) & 13.9507(6) & 39.270(6) & $06^{\rm h}\, 12^{\rm m}\, 44\, \fs087$ & $37\arcdeg\, 21\arcmin\, 40\, \farcs2$ & 175.44220 & 9.08013 \\
  J0738$+$6904 & 102.51871(7) & 46.69934(9) & 17.22(2) & $07^{\rm h}\, 38^{\rm m}\, 022\, \fs61$ & $69\arcdeg\, 04\arcmin\, 20\, \farcs1$ & 146.59345 & 29.37720 \\
  J0740$+$6620 & 103.7591384(1) & 44.1025059(1) & 14.92(2) & $07^{\rm h}\, 40^{\rm m}\, 45\, \fs799$ & $66\arcdeg\, 20\arcmin\, 33\, \farcs6$ & 149.72969 & 29.59937 \\
  J0747$+$6646 & 104.53950(2) & 44.69663(5) & 27.576(3) & $07^{\rm h}\, 47^{\rm m}\, 39\, \fs689$ & $66\arcdeg\, 46\arcmin\, 56\, \farcs8$ & 149.21764 & 30.28271 \\
  J0944$+$4106 & 134.006413(9) & 25.83992(3) & 21.41(3) & $09^{\rm h}\, 44^{\rm m}\, 18\, \fs141$ & $41\arcdeg\, 06\arcmin\, 04\, \farcs6$ & 180.43732 & 49.37512 \\
  J1059$+$6459 & 131.4687(7) & 51.9583(6) & 18.5(4) & $10^{\rm h}\, 59^{\rm m}\, 27\, \fs511$ & $64\arcdeg\, 59\arcmin\, 31\, \farcs8$ & 140.25104 & 48.19754 \\
  J1125$+$7819 & 115.6292886(9) & 62.4520225(4) & 11.219201 & $11^{\rm h}\, 25^{\rm m}\, 59\, \fs851$ & $78\arcdeg\, 19\arcmin\, 48\, \farcs7$ & 128.28875 & 37.89467 \\
  J1624$+$8643 & 93.79282(3) & 69.51940(1) & 46.43(2) & $16^{\rm h}\, 24^{\rm m}\, 032\, \fs75$ & $86\arcdeg\, 43\arcmin\, 13\, \farcs2$ & 119.92489 & 29.05717 \\
  J1628$+$4406 & 229.913199(1) & 64.4120718(2) & 7.32981(2) & $16^{\rm h}\, 28^{\rm m}\, 50\, \fs313$ & $44\arcdeg\, 06\arcmin\, 42\, \farcs6$ & 69.23851 & 43.61575 \\
  J1641$+$8049 & 101.8728674(9) & 74.8941763(2) & 31.08960(3) & $16^{\rm h}\, 41^{\rm m}\, 20\, \fs843$ & $80\arcdeg\, 49\arcmin\, 52\, \farcs9$ & 113.84003 & 31.76257 \\
  J1647$+$6608 & 174.9821(4) & 82.74687(3) & 22.55(7) & $16^{\rm h}\, 47^{\rm m}\, 32\, \fs522$ & $66\arcdeg\, 08\arcmin\, 22\, \farcs2$ & 97.18025 & 37.03415 \\
  J1710$+$4923 & 243.4177674(2) & 71.67561410(7) & 7.08493(2) & $17^{\rm h}\, 10^{\rm m}\, 04\, \fs442$ & $49\arcdeg\, 23\arcmin\, 11\, \farcs4$ & 75.92934 & 36.44891 \\
  J1800$+$5034 & 270.4263(1) & 74.01172(4) & 22.71(6) & $18^{\rm h}\, 00^{\rm m}\, 44\, \fs372$ & $50\arcdeg\, 34\arcmin\, 21\, \farcs7$ & 78.12962 & 28.43981 \\
  J1815$+$5546 & 281.2542(7) & 79.0649(2) & 58.999(7) & $18^{\rm h}\, 15^{\rm m}\, 05\, \fs739$ & $55\arcdeg\, 46\arcmin\, 23\, \farcs2$ & 84.31588 & 27.13337 \\
  J1821$+$4147 & 279.69363(3) & 65.03943(1) & 40.673(3) & $18^{\rm h}\, 21^{\rm m}\, 52\, \fs346$ & $41\arcdeg\, 47\arcmin\, 02\, \farcs5$ & 69.53713 & 22.90515 \\
  J1859$+$7654 & 72.6850(2) & 78.72083(4) & 47.25(7) & $18^{\rm h}\, 59^{\rm m}\, 36\, \fs019$ & $76\arcdeg\, 54\arcmin\, 55\, \farcs6$ & 108.36384 & 25.91076 \\
  J1923$+$4243 & 305.93303(4) & 63.58758(4) & 52.99(5) & $19^{\rm h}\, 23^{\rm m}\, 015\, \fs24$ & $42\arcdeg\, 43\arcmin\, 18\, \farcs6$ & 74.71528 & 12.64331 \\
  J1934$+$5219 & 320.8887(1) & 71.6218(1) & 71.9(1) & $19^{\rm h}\, 34^{\rm m}\, 23\, \fs892$ & $52\arcdeg\, 19\arcmin\, 57\, \farcs4$ & 84.49263 & 15.04604 \\
  J1938$+$6604 & 8.602915(1) & 80.1174609(2) & 41.2427(1) & $19^{\rm h}\, 38^{\rm m}\, 56\, \fs919$ & $66\arcdeg\, 04\arcmin\, 31\, \farcs7$ & 97.94564 & 20.03050 \\
  J1941$+$4320 & 313.54787(3) & 62.971984(8) & 79.361(8) & $19^{\rm h}\, 41^{\rm m}\, 58\, \fs915$ & $43\arcdeg\, 20\arcmin\, 06\, \farcs3$ & 76.84556 & 9.86437 \\
  J1942$+$8106 & 75.79591(5) & 74.12426(1) & 40.24(3) & $19^{\rm h}\, 42^{\rm m}\, 54\, \fs669$ & $81\arcdeg\, 06\arcmin\, 17\, \farcs3$ & 113.37123 & 24.82517 \\
  J1954$+$4357 & 318.69172(4) & 62.63881(2) & 130.30(5) & $19^{\rm h}\, 54^{\rm m}\, 038\, \fs45$ & $43\arcdeg\, 57\arcmin\, 37\, \farcs3$ & 78.52732 & 8.16863 \\
  J1955$+$6708 & 16.178385(1) & 78.7214729(2) & 57.1478(1) & $19^{\rm h}\, 55^{\rm m}\, 38\, \fs764$ & $67\arcdeg\, 08\arcmin\, 15\, \farcs1$ & 99.68213 & 18.93834 \\
  J2001$+$4258 & 320.10684(2) & 61.24332(2) & 54.93(3) & $20^{\rm h}\, 01^{\rm m}\, 010\, \fs59$ & $42\arcdeg\, 58\arcmin\, 06\, \farcs2$ & 78.27728 & 6.64261 \\
  J2017$+$5906 & 350.7866(2) & 72.89756(3) & 60.28(6) & $20^{\rm h}\, 17^{\rm m}\, 44\, \fs555$ & $59\arcdeg\, 06\arcmin\, 46\, \farcs9$ & 93.60825 & 12.90340 \\
  J2027$+$7502 & 51.5269(2) & 75.59357(2) & 11.71(1) & $20^{\rm h}\, 27^{\rm m}\, 23\, \fs274$ & $75\arcdeg\, 02\arcmin\, 29\, \farcs2$ & 108.36769 & 20.31252 \\
  J2123$+$5434 & 358.500835(3) & 63.277719(2) & 31.760(3) & $21^{\rm h}\, 23^{\rm m}\, 21\, \fs681$ & $54\arcdeg\, 34\arcmin\, 07\, \farcs7$ & 95.69847 & 3.09706 \\
  J2137$+$6428 & 20.1556(4) & 68.10698(6) & 106.0(3) & $21^{\rm h}\, 37^{\rm m}\, 20\, \fs257$ & $64\arcdeg\, 28\arcmin\, 41\, \farcs7$ & 103.85139 & 9.06360 \\
  J2208$+$4056 & 354.46060(4) & 47.91605(4) & 11.837(9) & $22^{\rm h}\, 08^{\rm m}\, 01\, \fs991$ & $40\arcdeg\, 56\arcmin\, 01\, \farcs8$ & 92.57410 & -12.11198 \\
  J2228$+$6447 & 27.9436(2) & 63.61524(4) & 193.6(2) & $22^{\rm h}\, 28^{\rm m}\, 40\, \fs502$ & $64\arcdeg\, 47\arcmin\, 19\, \farcs4$ & 108.46152 & 6.00724 \\
  J2241$+$6941 & 39.1473(1) & 65.08555(2) & 67.67(7) & $22^{\rm h}\, 41^{\rm m}\, 20\, \fs295$ & $69\arcdeg\, 41\arcmin\, 59\, \farcs7$ & 112.02560 & 9.63205 \\
  J2310$+$6706 & 37.3837(3) & 61.4356(4) & 97.7(2) & $23^{\rm h}\, 10^{\rm m}\, 42\, \fs077$ & $67\arcdeg\, 06\arcmin\, 52\, \farcs1$ & 113.35114 & 6.13557 \\
  J2312$+$6931 & 41.8523(2) & 62.62711(4) & 71.6(1) & $23^{\rm h}\, 12^{\rm m}\, 038\, \fs93$ & $69\arcdeg\, 31\arcmin\, 04\, \farcs0$ & 114.43546 & 8.29219 \\
  J2351$+$8533 & 78.8764(3) & 66.3376(2) & 38.5(4) & $23^{\rm h}\, 51^{\rm m}\, 03\, \fs261$ & $85\arcdeg\, 33\arcmin\, 20\, \farcs7$ & 121.67724 & 22.83186 \\
  \enddata
  \tablecomments{Ecliptic coordinates use the IERS2010 value of the
  obliquity of the ecliptic referenced to J2000 \citep{cwc03}.  Values
  in parentheses are the $1$-$\sigma$ uncertainty in the last digit as
  reported by \texttt{TEMPO}.}
\end{deluxetable*}

\begin{deluxetable*}{lcccc}[t!]
  \tabletypesize{\footnotesize}
  \tablewidth{\textwidth}
  \tablecaption{ELL1 Binary Parameters of GBNCC
    Pulsars \label{table:binary}}
  \tablecolumns{5}
  \tablehead{\colhead{Parameter} & \colhead{J0740$+$6620} & \colhead{J1125$+$7819} & \colhead{J1641$+$8049} & \colhead{J1938$+$6604}}
  \startdata
  \cutinhead{Measured Parameters}
  $P_{\rm B}$ (d) & 4.766944616(3) & 15.35544590(2) & 0.0908739634(1) & 2.467162727(1) \\ 
  $a\sin{i}/c$ ($\s$) & 3.977556(1) & 12.1924288(7) & 0.0640793(3) & 8.950738(1) \\ 
  $T_{\rm asc}$ (MJD) & 56155.3684710(2) & 56157.4763453(3) & 56220.5100737(1) & 56366.0967005(1) \\ 
  $\epsilon_1$ & $-$5.6(4)$\times$10$^{-6}$ & $-$1.28(1)$\times$10$^{-5}$ & 0.000111(8) & 4.8(3)$\times$10$^{-6}$ \\ 
  $\epsilon_2$ & $-$2.0(2)$\times$10$^{-6}$ & 1.0(1)$\times$10$^{-6}$ & $-$5.4(6)$\times$10$^{-5}$ & $-$2.81(3)$\times$10$^{-5}$ \\ 
  \cutinhead{Derived Parameters}
  $T_0$ (MJD) & 56156.30(3) & 56153.82(2) & 56220.4939(8) & 56366.030(5) \\ 
  $e$ & 5.9(4)$\times$10$^{-6}$ & 1.29(1)$\times$10$^{-5}$ & 0.000123(8) & 2.85(3)$\times$10$^{-5}$ \\ 
  $\omega$ ($\arcdeg$) & 1.23(4) & $-$1.495(8) & $-$1.12(5) & $-$0.17(1) \\ 
  $f_{\rm M}$ ($\Msun$) & 0.002973387(3) & 0.008253317(1) & 3.42102(5)$\times$10$^{-5}$ & 0.12649217(6) \\ 
  $M_{\rm c,min}$ ($\Msun$) & 0.2 & 0.29 & 0.04 & 0.87 \\ 
  \enddata
  \tablecomments{All timing models presented here use the ELL1 binary
  model, which is appropriate for low-eccentricity orbits.  Binary
  parameters for the relativistic binary PSR J0509+3801 are shown in
  Table \ref{table:J0509}.  Values in parentheses are the
  $1$-$\sigma$ uncertainty in the last digit as reported by
  \texttt{TEMPO}.}
\end{deluxetable*}
\begin{deluxetable*}{lccccccccc}[b!]
  \centering
  \tabletypesize{\footnotesize}
  \tablewidth{0pt}
  \tablecolumns{10}
  \tablecaption{Proper Motions and Kinematic Corrections for Five
    GBNCC Pulsars \label{table:pms}}
  \tablehead{
    \colhead{PSR}                                    &
    \colhead{$\mu_\lambda$}     &
    \colhead{$\mu_\beta$}       &
    \colhead{$D_{\rm DM}$} &
    \colhead{$v_{\rm t}$}            &
    \colhead{$\dot{P}_{\rm G}$}    &
    \colhead{$\dot{P}_{\rm S}$}    &
    \colhead{$B_{\rm surf}$} &
    \colhead{$\tau_{\rm c}$} &
    \colhead{$\dot{E}$} \\
    \colhead{}                      &
    \colhead{($\mathrm{mas}\, \yr^{-1}$)} &
    \colhead{($\mathrm{mas}\, \yr^{-1}$)} &
    \colhead{(\kpc)} &
    \colhead{($\km\, \s^{-1}$)}      &
    \colhead{($10^{-21}$)}           &
    \colhead{($10^{-21}$)}           &
    \colhead{($10^8\; \gauss$)} &
    \colhead{(Gyr)} &
    \colhead{($10^{33}\; \erg\, \s^{-1}$)}
  }
  \startdata
  J0740+6620 & $-3.2(4)$  & $-33.7(6)$ & 0.7\tablenotemark{a} & 112\tablenotemark{a} & $-0.14$\tablenotemark{a} & 5.6\tablenotemark{a} & 1.4\tablenotemark{a} & 6.8\tablenotemark{a} & 11\tablenotemark{a} \\
             &            &            & 0.9\tablenotemark{b} & 144\tablenotemark{b} & $-0.13$\tablenotemark{b}  & 7.2\tablenotemark{b} & 1.2\tablenotemark{b} & 8.9\tablenotemark{b} & 8.4\tablenotemark{b} \\
  J1125+7819 & \phantom{$-$}$20(3)$ & \phantom{$-$}$25(3)$ & 0.6\tablenotemark{a} & 91\tablenotemark{a}  & -0.40\tablenotemark{a} & 6.3\tablenotemark{a} & 0.77\tablenotemark{a} & 48\tablenotemark{a} & 0.73\tablenotemark{a} \\
             &                      &                      & 0.8\tablenotemark{b} & 121\tablenotemark{b} & -0.44\tablenotemark{b} & 8.4\tablenotemark{b} & \nodata & \nodata & \nodata \\
  J1641+8049 & $-11(1)$  & \phantom{$-$}$37(3)$ & 1.7\tablenotemark{a} & 311\tablenotemark{a} & -0.32\tablenotemark{a} & 12\tablenotemark{a}  & \nodata & \nodata & \nodata \\
             &           &                      & 2.1\tablenotemark{b} & 384\tablenotemark{b} & -0.35\tablenotemark{b} & 15\tablenotemark{b}  & \nodata & \nodata & \nodata \\
  J1710+4923 & $-50.4(2)$ & $-44.7(2)$ & 0.7\tablenotemark{a} & 224\tablenotemark{a} & -0.41\tablenotemark{a} & 25\tablenotemark{a}  & \nodata & \nodata & \nodata\\
             &            &            & 0.5\tablenotemark{b} & 160\tablenotemark{b} & -0.34\tablenotemark{b} & 18\tablenotemark{b}  & 0.43\tablenotemark{b} & 92\tablenotemark{b} & 0.65\tablenotemark{b} \\
  J1955+6708 & $-3(1)$ & \phantom{$-$}$10(2)$ & 3.4\tablenotemark{a}  & 168\tablenotemark{a} & -2.2\tablenotemark{a} & 2.9\tablenotemark{a} &  3.2\tablenotemark{a} & 11\tablenotemark{a} & 0.75\tablenotemark{a} \\
             &         &                      & 10.1\tablenotemark{b} & 500\tablenotemark{b} & -2.9\tablenotemark{b} & 8.6\tablenotemark{b} &  2.5\tablenotemark{b} & 20\tablenotemark{b} & 0.43\tablenotemark{b} \\
  \enddata
  \tablenotetext{a}{These values use the NE2001 \citep{cl02} DM-inferred distance.}
  \tablenotetext{a}{These values use the YMW16 \citep{ymw17} DM-inferred distance.}
  \tablecomments{Values of $B_{\rm surf}$, $\tau_{\rm c}$, and
    $\dot{E}$ that have been corrected for the Shklovskii effect are
    shown for pulsars where $\dot{P}_{\rm S} < \dot{P}_{\rm obs}$.}
\end{deluxetable*}
\FloatBarrier

\subsection{Proper Motions and Kinematic Corrections}
\label{sec:pms}

We have measured timing proper motions for five MSPs: PSRs J0740+6620,
J1125+7819, J1641+8049, J1710+4923, and J1955+6708.  Table
\ref{table:pms} lists the measured proper motions, estimated
transverse velocity, $v_{\rm t}$ calculated using the DM-inferred
distances in both the NE2001 \cite{cl02} and YMW16 \citep{ymw17}
models, and kinematic corrections.

The observed pulsar spin down is a contribution of several effects:
\begin{equation}
  \dot{P}_{\rm obs} = \dot{P}_{\rm int} + \dot{P}_{\rm G} + \dot{P}_{\rm S}
\end{equation}
where $\dot{P}_{\rm G}$ is caused by acceleration between the pulsar
and Solar System Barycenter in a differential Galactic potential and
$\dot{P}_{\rm s}$ is the Shklovskii effect for pulsars.  Following
\citet{nt95}, the bias arising from acceleration perpdendicular to the
Galactic plane ($a_z$) is given by
\begin{equation}
  \frac{\dot{P}_{\rm G,\perp}}{P} = \frac{a_z \sin{b}}{c}
\end{equation}
where $b$ is the Galactic latitude and $a_z$ is
\begin{equation}
  \frac{a_z}{c} = -1.09 \times 10^{-19} \s^{-1} \left[ \frac{1.25
      z}{(z^2 + 0.0324)^{1/2}} + 0.58 z \right] ,
\end{equation}
where $z = (D/\kpc)\, \sin{b}$.  The planar component is given by
\begin{equation}
  \frac{\dot{P}_{\rm G,p}}{P} = -\cos{b}
  \left(\frac{\Theta_0^2}{c\, R_0} \right) \left(\cos{\ell} +
  \frac{\beta}{\sin^2{\ell} + \beta^2} \right)
\end{equation}
where $\Theta_0 = 240\; \km\, \s^{-1}$ is the Sun's galactocentric
velocity, $R_0 = 8.34\; \kpc$ is the Sun's galactocentric distance,
and $\beta = (D_{\rm DM}/R_0)\; \cos{b} - \cos{\ell}$, and $\ell$ is
the Galactic longitude. The Shklovskii effect \citep{shk70} is
\begin{equation}
  \frac{\dot{P}_{\rm S}}{P} = 2.43 \times 10^{-21}\; \s^{-1}
  \left( \frac{D_{\rm DM}}{\kpc} \right ) \left(\frac{\mu}{10^{-3}\;
    \arcsec\; \yr^{-1}} \right )^2,
\end{equation}

Note that in the case of PSRs J1125+7819, J1641+8049, and J1710+4923,
the DM-inferred distance under one or both of the NE2001 and YMW16
models leads to values of $\dot{P}_{\rm S}$ larger than the observed
$\dot{P}$.  None of these pulsars show any evidence that they are
being spun up via accretion, so in all three cases $\dot{P}_{\rm int}$
should be positive.  Since $|\dot{P}_{\rm G}| \ll \dot{P}_{\rm S}$ in
these cases, we can ignore the Galactic acceleration component and
enforce $\dot{P}_{\rm S} < \dot{P}_{\rm obs}$ thereby setting an upper
limit on the distances (note that we do not use these distance limits
to re-estimate $\dot{P}_{\rm G}$).  For PSR J1125+7819, $D < 0.7\;
\kpc$ and $v_{\rm t} < 104\; \km \, \s^{-1}$; PSR J1641+8049, $D <
1.2\; \kpc$ and $v_{\rm t} < 230\; \km\, \s^{-1}$; for PSR J1710+4923,
$D < 0.5\; \kpc$ and $v_{\rm t} < 160\; \km, \s^{-1}$.

The mean transverse velocity of our sample of MSPs is $\mu_v = 152\;
\km\, \s^{-1}$, with a standard deviation of $\sigma_v = 48\; \km\,
\s^{-1}$ using the NE2001 DM-inferred distances, and $\mu_v = 234\;
\km\, \s^{-1}$, $\sigma_v = 143\; \km\, \s^{-1}$ using the YMW16
DM-inferred distances (here we use the upper limits on $v_{\rm t}$ as
appropriate).  Uncertainty in the DM-inferred distances makes a
precise measurement of transverse velocity difficult.  With this
caveat in mind, the velocities that we estimate are somewhat higher
than those found by other authors, though still within one to two
standard deviations: \citet{tsb+99} find $\mu_v = 85\; \km\, \s^{-1},
\sigma_v = 13\; \km\, \s^{-1}$ for a sample of 13 MSPs; \citet{hllk05}
find $\mu_v = 87\; \km\, \s^{-1}, \sigma_v = 13\; \km\, \s^{-1}$ for a
sample of 35 MSPs; and \citet{gsf+11} find $\mu_v = 88\; \km\,
\s^{-1}, \sigma_v = 12\; \km\, \s^{-1}$ for a sample of five MSPs.

\begin{deluxetable}{lr}[h!]
  \tabletypesize{\footnotesize}
  \tablewidth{0.8\columnwidth}
  \tablecaption{Binary Parameters of PSR J0509+3801 \label{table:J0509}}
  \tablecolumns{2}
  \tablehead{\colhead{Parameter}& \colhead{Value}}
  \startdata
  \cutinhead{Measured Parameters}
  $P_{\rm b}$ (d)  \hphantom{xxxxxxxxxxxxxxxxxxxxxx} & 0.379583785(3)    \\
  $a \sin{i}/c$ ($\s$)                     & 2.0506(3)    \\
  $T_0$ (MJD)                                & 56075.412714(3)    \\
  $e$                                        & 0.586400(6)     \\
  $\omega$ ($\arcdeg$)                       & 127.77(1)    \\
  $M_{\rm tot}$ (\Msun)                    & 2.805(3)  \\
  $M_{\rm c}$ (\Msun)                      & 1.46(8)    \\
  \cutinhead{Derived Parameters}
  $M_{\rm p}$ (\Msun)                      & 1.34(8)    \\ 
  $f_{\rm M}$ (\Msun)                      & 0.06425(3)    \\
  $\dot{\omega}$ ($\arcdeg\; \yr^{-1}$)  & 3.031 \\
  $\gamma$ ($\s$)                            & 0.0046 \\
  $\dot{P}_{\rm b}$ ($10^{-12}$)       & $-1.39$ \\
  $\sin{i}$                                & 0.55 \\
  \enddata
  \tablecomments{Values in parentheses are the $1$-$\sigma$
  uncertainty in the last digit as reported by \texttt{TEMPO}.}
\end{deluxetable}

\subsection{Discussions of Individual Systems}
\label{sec:systems}

\subsubsection{PSR J0509+3801: A New Double Neutron Star System}
\label{sec:J0509}

\begin{figure}[t!]
  \centering \includegraphics[width=\columnwidth]{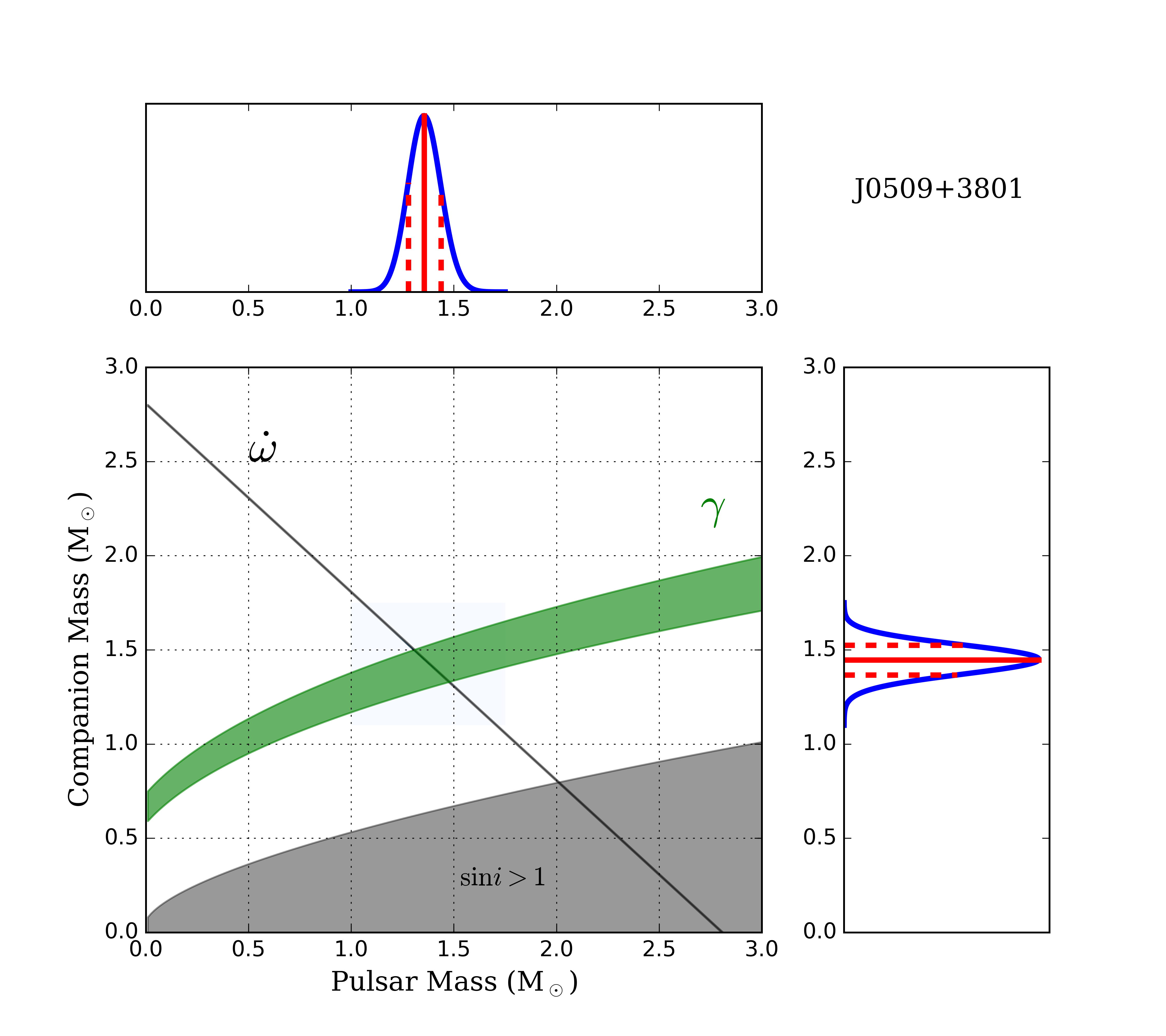}
  \caption{Estimates of the posterior probability density and
    marginalized distribution functions (blue-solid curves) for mass
    estimates of the PSR J0509+3801 system. The regions of allowed
    masses for individual PK parameters are shown as color-shaded
    curves in the density map; the thickness of each color-shaded
    curve represents the 1$\sigma$ uncertainty determined by
    \texttt{TEMPO} (note that the uncertainties on $\dot{\omega}$ are
    too small to be visible on this scale). The red-solid lines are
    the median values for each component mass, and the red-dashed
    lines represent the edges of the 68.3\% credible
    intervals. \label{fig:J0509_M-M}}
\end{figure}
 
PSR J0509+3801 is part of a highly eccentric binary system with
$P_{\rm b} = 9.11\; \hr$ and $e = 0.586$.  Early in our timing
campaign we measured a significant change in the longitude of
periastron, $\dot{\omega} = 3.031(2)\arcdeg\; \yr^{-1}$.  In general
relativity (GR), $\dot{\omega}$ is related to the total system mass
and our measured value implies $M_{\rm tot} = 2.81\;
\Msun$. Additional timing observations resulted in the measurement of
the amplitude of the Einstein delay due to gravitational redshift and
time dilation, $\gamma = 0.0046(3)\; \s$.  With the measurement of two
post-Keplerian parameters, we were able to measure the masses of both
PSR J0509+3801 and its companion within the framework of GR.  Using
the DDGR timing model, which uses the masses as free parameters, we
find $M_{\rm p} = 1.34(8)\; \Msun$ and $M_{\rm c} = 1.46(8)\; \Msun$.
Table \ref{table:J0509} gives our complete timing solution.  The high
companion mass and eccentricity lead us to classify this as a new DNS
system.

We also performed a separate Bayesian analysis of the masses.  The
two-dimensional probability map was computed using a $\chi^2$-grid
method and the best-fit timing solution, where the Shapiro-delay
parameters are held fixed at each mass-mass coordinate while all other
parameters are allowed to float freely when using \texttt{TEMPO} and
the current TOA data set. We then used the procedure outlined by
\citet{sna+02} to compute probability densities from the
$\chi^2$-grid, and then marginalized over each mass coordinate to
obtain one-dimensional probability distribution functions (PDFs) for
the pulsar and companion masses. We finally computed the equal-tailed,
68.3\% credible intervals and median values of the neutron-star masses
from these PDFs (see Figure \ref{fig:J0509_M-M}).  From this analysis
we obtain estimates of $M_{\rm p} = 1.36(8) \; \Msun$ and $M_{\rm c} =
1.45(8)\; \Msun$. These estimates and the credible intervals are
consistent with uncertainties determined by the least-squares fit
obtained from \texttt{TEMPO}.

The Shapiro delay cannot be measured in this system with the current
data presented in this work, since the RMS timing residual for
J0509+3801 exceeds typical amplitudes of the relativistic
signal. However, the neutron-star masses are estimated using two PK
measurements under the assumption that GR is correct, and without any
consideration of the binary mass function. We therefore use the mass
function and the two mass constraints to estimate the inclination
angle, and find that $i = 33^{+2}_{-2}$ degrees or $i = 147^{+2}_{-2}$
degrees. The two inclination estimates are allowed since the mass
function depends on $\sin i$ only, and therefore yields no constraint
on the sign of $\cos i$.

The masses of both stars and their mass ratio are similar to those in
most other DNS systems \citep{kkdt13}.

\subsubsection{PSR J0740+6620: Constraints on Pulsar and Companion Mass}
\label{sec:J0740}

We report here a weak, $2$-$\sigma$ measurement of Shapiro delay in
PSR J0740+6620.  We find best-fit values of the Shapiro range and
shape parameters of $r = 1.0(3)\; \us$ and $s = 1.00017(99)$ when
using the ELL1 model, which is a theory-independent model.  Within the
framework of GR the Shapiro parameters become $r = T_\sun M_{\rm c}$,
implying $M_{\rm c} = 0.21(6)$, and $s = \sin{i}$.  The best-fit value
of $s$ is $1$-$\sigma$ consistent with $i < 90\arcdeg$ but includes a
non-physical range.  A dedicated campaign to observe PSR J0740+6620
near conjunction, when Shapiro delay is maximum, will be the subject
of a future study, but our current results are consistent with a
nearly edge-on system and with a $\sim 0.2\; \Msun$ companion.

We can constrain the companion mass along independent lines of
reasoning, as well.  The low eccentricity and few-day orbital period
of PSR J0740+6620 are consistent with expectations for a He WD
companion.  A well-defined relationship is observed between $P_{\rm
  b}$ and $M_{\rm c}$ in such systems \citep{ts99,imt+16}, and in the
case of PSR J0740+6620 predicts a companion mass $\sim 0.2\; \Msun$.
We can also calculate the minimum companion mass by assuming $i =
90\arcdeg$ and $M_{\rm p} = 1.4\; \Msun$, and find $M_{\rm c, min} =
0.2\; \Msun$.  Both of these values are consistent with the tentative
measurement of $r$ and with an inclination angle close to $90\arcdeg$,
for a pulsar mass of $1.4\; \Msun$.  We therefore conclude that the
pulsar's companion is a He WD and that the pulsar mass is close to the
canonical value. 

\subsubsection{PSR J1938+6604: An Intermediate Mass Binary Pulsar}
\label{sec:J1938}

PSR J1938+6604 is a partially recycled binary pulsar with $P = 22\;
\ms$.  The minimum companion mass assuming $M_{\rm p} = 1.4\; \Msun$
and $i = 90\arcdeg$ is $0.87\; \Msun$.  Motivated by the potentially
high companion mass, we conducted a campaign to measure Shapiro delay
in PSR J1938+6604 using the GBT.  We observed the pulsar for six hours
around conjunction and for two to four hours at other select orbital
phases where the measurable Shapiro delay signature is predicted to be
at a local maximum.  Our campaign totaled 21 hours and was conducted
at a center frequency of $1.4\; \GHz$ using coherent dedispersion (see
Table \ref{table:obs} for details).  However, we were unable to detect
Shapiro delay.

Our non-detection implies that the system is not highly inclined ($i
\lesssim 80\arcdeg$), and if we assume $M_{\rm p} = 1.4\; \Msun$, the
companion mass limit is $M_{\rm c} > 0.88\; \Msun$.  We see no
evidence for eclipses, variations in DM, or changes in orbital period,
and we find no optical companion in the Digital Sky Survey or
(Pan-STARRS) archives, so the companion is unlikely to be a main
sequence or giant star.  The eccentricity of the system is measurable
and very low ($e = 2.8 \times 10^{-5}$) unlike double neutron star
systems.  PSR J1938+6620 is therefore most likely an intermediate mass
binary pulsar (IMBP; \citealt{clm+01}) with a CO/ONeMg WD companion.

\subsubsection{PSR J1641+8049: A Pulsar in a Black Widow Binary System}
\label{sec:J1641}

PSR J1641+8049 is a fast MSP that exhibits eclipses and has a very low
mass companion ($M_{\rm c, min} = 0.04\; \Msun$), making it a member
of the black widow class of binary pulsars.  In black widow systems an
energetic MSP ablates its companion, forming a low-mass remnant, and
perhaps eventually an isolated MSP \citep{fst88,pebk88}.  Although we
never observed ingress and egress for the same eclipse, based on
pulsar timing the eclipses lasted from orbital phases $\approx
0.205$--$0.355$ at $350\; \MHz$, or about 20 minutes.  TOAs affected
by excess DM near ingress and egress were not included in our timing
analysis.

Since black widows have high spin-down luminosities, they are commonly
detected in gamma-rays \citep[e.g.][]{rrc+11,wkh+12,egc+12,ckr+15}.
The upper limit on $\dot{E}$ for PSR J1641+8049 is in the middle of
the sample of 51 publically listed \textit{Fermi}-detected
MSPs\footnote{\url{https://confluence.slac.stanford.edu/display/GLAMCOG/Public+List+of+LAT-Detected+Gamma-Ray+Pulsars}}
with measured $\dot{E}$ and distances listed in the ATNF pulsar
catalog.  We searched for pulsations in \textit{Fermi} Large Area
Telescope (LAT) events by downloading all events around a $3\arcdeg$
region of interest centered on the timing position reported in Table
\ref{table:astrometric_dm}, recorded between MJDs 54802.65 and
57857.5\footnote{Mission Elapsed Times 249925417--513863184}, and in
an energy range of $200$-$30000\; \MeV$.  We used the
\texttt{fermiphase} routine of the
\texttt{PINT}\footnote{\url{http://nanograv-pint.readthedocs.io/en/latest/}}
pulsar timing and data analysis package to read the \textit{Fermi}
events, compute a pulse phase for each event, and calculate the
$H$-statistic \citep{drs89} of the resulting light curve.  This
results in $H = 3.26$, corresponding to an equivalent Gaussian $\sigma
= 1.10$, i.e. no significant pulsations are detected.  It is plausible
that the Shklovskii correction for PSR J1641+8049 is large and that
the true $\dot{E}$ is much lower than the nominal value, explaining
our non-detection.

\subsubsection{Optical Observations of the PSR J1641+8049 System}
\label{sec:J1641_optical}

We detected a faint optical counterpart to PSR J1641+8049 by examining
the stacked images of the Panoramic Survey Telescope and Rapid
Response System (PanSTARRS) $3\pi$ survey data release 1 (PS1;
\citealt{cmm+16}).  It was only visible in the $r$ and $i$ bands, and
was not listed in the photometric catalogs.  Therefore, we determined
rough photometry ourselves using the images and \texttt{sextractor}
\citep{ba96}, finding $r=24.0\pm0.3$ and $i=23.2\pm0.2$.

\begin{figure}[t]
  \centering
  \includegraphics[width=\columnwidth]{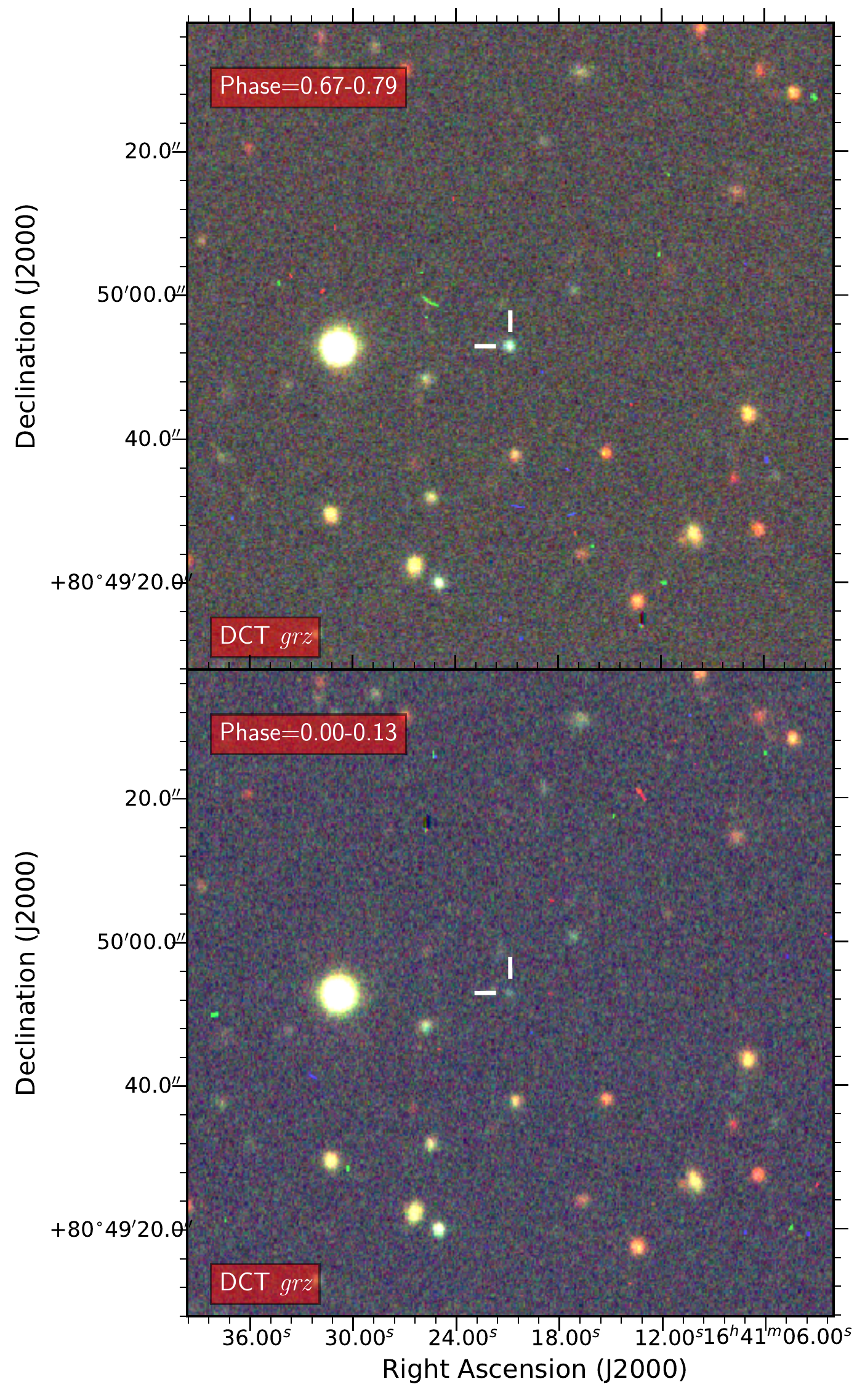} \\
  \caption{Composite rgb images of PSR J1641+8049, taken with the
    DCT's LMI at two different orbital phases: 0.67--0.79 (close to
    photometric maximum) at the top, and 0.00-0.13 (close to
    photometric minimum) at the bottom.  The composites are made from
    the $g$, $r$, and $z$ filters.  The position of PSR J1641+8049 is
    indicated with the ticks toward the center: the astrometric
    uncertainties are dominated by the astrometric calibration of the
    optical images, which has uncertainties of roughly $0\farcs2$.
    The image are $1.5\arcmin$ on a side, with north up and east to
    the left.
    \label{fig:J1641_optical}}
\end{figure}

To improve the photometry we observed the field with the 4.3-m
Discovery Channel Telescope (DCT) in Happy Jack, Arizona, using the
Large Monolith Imager (LMI).  Five 300-s observations in each of $g$,
$r$, $i$, and $z$ filters ($487\; \nm$, $622\; \nm$, $755\; \nm$, and
$868\; \nm$, respectively) were taken on 2017 March 16 from 08:58 UTC
to 10:42 UTC, spanning almost $2\; \hr$, or slightly less than one
orbit.  Standard CCD reduction techniques (e.g., bias subtraction,
flat fielding) were applied using a custom IRAF pipeline.  Individual
exposures were astrometrically algined with respect to reference stars
from the Sloan Digital Sky Survey \citep{aaa+14} using \texttt{SCAMP}
\citep{bert06}.  Composite rgb images near photometric maximum and
minimum are shown in Figure \ref{fig:J1641_optical}.  We calibrated
the photometry using \texttt{sextractor} and the PS1 catalog as a
reference, with roughly 40 unsaturated stars per image per filter.
The observation times were corrected to the solar system barycenter
using \texttt{PINT}.  We find a significant amount of photometric
variation over the course of one orbit, with $r$ ranging from $21.7$
to $>24.7$ and similar variability in the other filters.  We also
observed this system between March and April 2017 with the Sinistro
camera on the 1-m telescope at the McDonald Observatory.  There were
18 observations using the $r'$ filter and 16 using $i'$, all using
500-s exposures.  We detected the system on 2017 March 19 at an
orbital phase of $\phi_{\rm B} \approx 0.62$ (radio convention, where
the ascending node marks $\phi_{\rm B} = 0$) with $r' = 21.73 \pm
0.31$, and again on 2017 March 21 at $\phi_{\rm B} \approx 0.65$ with
$r' = 21.38 \pm 0.23$ and $i' = 21.57 \pm 0.46$.  All other
observations resulted in non-detections.

Based on positional agreement and strong photometric variability tied
to the orbital period, we can be certain that we have identified the
optical counterpart of PSR J1641+8049.  Future observations and
analysis will enable us to determine the range of radii and effective
temperatures for the companion and use to those to constrain the mass
of the pulsar and inclination of the binary system
\citep[e.g.][]{afw+13}.

\subsubsection{PSR J1628+4406: A New Mode Changing Pulsar}
\label{sec:J1628}

PSR J1628+4406 has two emission modes, distinguished by differences in
the amplitude of its pulse profile components (see Figure
\ref{fig:profiles}).  In the first mode, the main pulse has two
components with similar amplitudes (the leading component,
i.e. leftward in Figure \ref{fig:profiles}, is somewhat weaker than
the trailing component), and there is a strong interpulse separated by
$180\arcdeg$ in pulse phase from the trailer component of the main
pulse.  This general description of Mode 1 holds in both the $350\;
\MHz$ and $820\; \MHz$ bands.  In the $148\; \MHz$ band observed with
LOFAR, the leading component of the main pulse is blended with the
trailing component but, given the asymmetry in the profile, the
leading component seems to be much weaker.  In the second mode, the
amplitude of the leading component of the main pulse is greatly
reduced and the interpulse nearly vanishes, though it is still visible
when several observations in this mode are summed together.  Each
component remains at the same pulse phase in both modes.

\begin{figure}[t]
  \centering
  \includegraphics[width=\columnwidth]{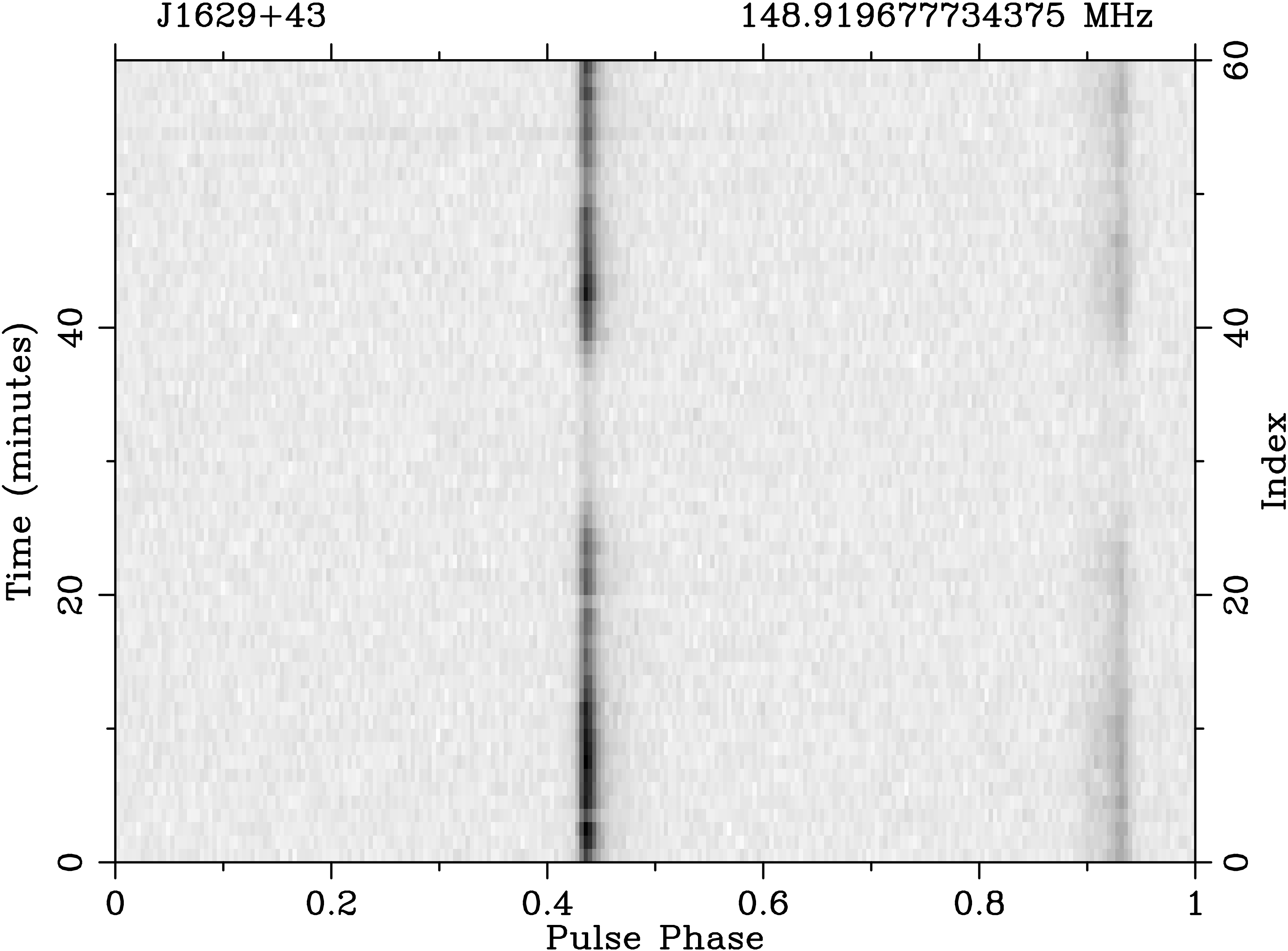}
  \caption{A phase-time plot of J1628+4406 from MJD 56639 as observed
    by LOFAR.  The pulsar experiences a significant drop in flux for
    15 minutes but does not appear to change modes.  No other similar
    events were observed.
    \label{fig:J1628_fading}}
\end{figure}

We observed PSR J1628+4406 for a total of $48091\; \s$ during our
timing campaign, split between the $148\; \MHz$ LOFAR band (where the
vast majority of the observations were made), and the $350\; \MHz$ and
$820\; \MHz$ bands of the GBT.  Of this, the pulsar spent $33156\; \s$
(70\%) in Mode 1 while the remaining time ($14574\; \s$; 30\%) was
spent in Mode 2.  We never witnessed a transition between modes during
an observation, but we did observe a single event in which the pulsar
significantly dropped in flux across both profile components for
approximately 15 minutes before recovering to its prior state (see
Fig. \ref{fig:J1628_fading}).  There is no record of an instrumental
failure that would account for this drop in flux, nor was there
indication of Solar activity that might cause ionospheric changes that
would impact the observation at this level, so the flux change would
appear to be a genuine phenomenon in this pulsar.

\subsubsection{PSR J2123+5434: A 138-ms Pulsar with a Low Magnetic Field}
\label{sec:J2122}

\begin{figure*}[t!]
\centering
\includegraphics[width=\textwidth]{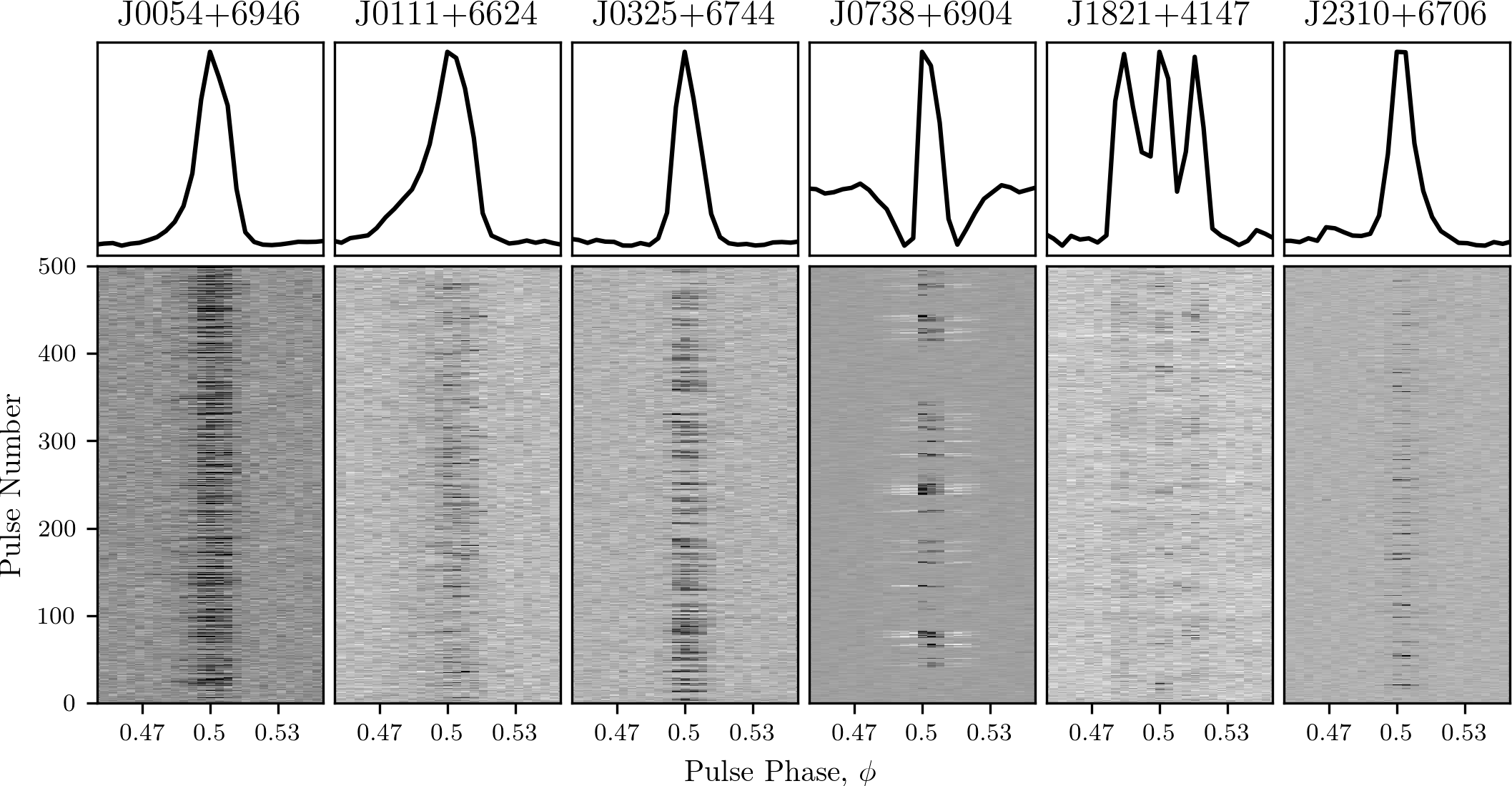}
\caption{Preliminary single pulse analysis for six GBNCC nulling
  pulsars with existing timing solutions. All data were taken
    at $820\; \MHz$.  For each source, 500 pulses are plotted in the
    bottom panels to show nulling behavior; folded profiles are shown
    in the top panels, corresponding to the same duration of pulse
    phase ($0.45<\phi<0.55$).  The apparent drop in flux on either
    side of the profile in PSR J0738+6904 is an artifact affecting
    subbanded data of bright pulsars.  Since it is clear when the
    pulsar is in an ON or OFF state, we do not expect it to bias the
    results of our nulling analysis. \label{fig:singlepulses}}
\end{figure*}

At the completion of our timing program, our timing solution for PSR
J2123+5434 did not constrain $\dot{P}$. Therefore, we obtained
additional observations for this pulsar through our survey program
about 3 years after the initial program, giving us a total time span
of 4 years. These additional observations enabled a $\dot{P}$
measurement of $1.7(1) \times 10^{-19}$, implying a surface magnetic
field of $B_{\rm surf} = 5.0(1) \times 10^9\; \gauss$. These values
are smaller than any known pulsar with a period $> 100\; \ms$ that has
a constrained $\dot{P}$ measurement. Figure \ref{fig:p-pdot} shows the
location of PSR J2123+5434 in the $P$-$\dot{P}$ plane in relation to
the pulsars in the ATNF pulsar catalog. The expected change in
$\dot{P}$ due to Galactic motion for PSR J2123+5434 (using the
DM-derived distance of $2.1\; \kpc$ \citep{cl02}) is about $-2.3
\times 10^{-20}$. After applying this correction, PSR J2123+5434
remains an outlier in comparison to the typical pulsar.  A potential
explanation for PSR J2123+5434's anomolously low $\dot{P}$ is that it
was partially recycled by a high mass companion star that later
underwent a supernova, disrupting the system. Alternatively, the
pulsar may be in a very wide binary, in which case we could be
observing orbital phases in which the pulsar is accelerating towards
the Earth, inducing a Doppler $\dot{P} < 0$ \citep[e.g. PSR
  J1024$-$0719][]{dkn+16,bjs+16}. We examined archival infrared and
optical data at the timing position of J2123+5434, but did not
identify a counterpart at any wavelength.

\section{Nulling and Intermittent Pulsars}
\label{sec:nullers}

Given the rarity of the phenomenon, nulling studies typically
characterize pulsars by their ``nulling fractions" (NFs; the fraction
of time spent in a null state) and look for correlations between NF
and other measured/derived parameters like spin period, characteristic
age, and profile morphology
\citep[e.g.][]{rit76,ran86,big92,viv95,wmj07}.

\begin{figure*}[t!]
  \centering
  \includegraphics[height=0.7\textwidth]{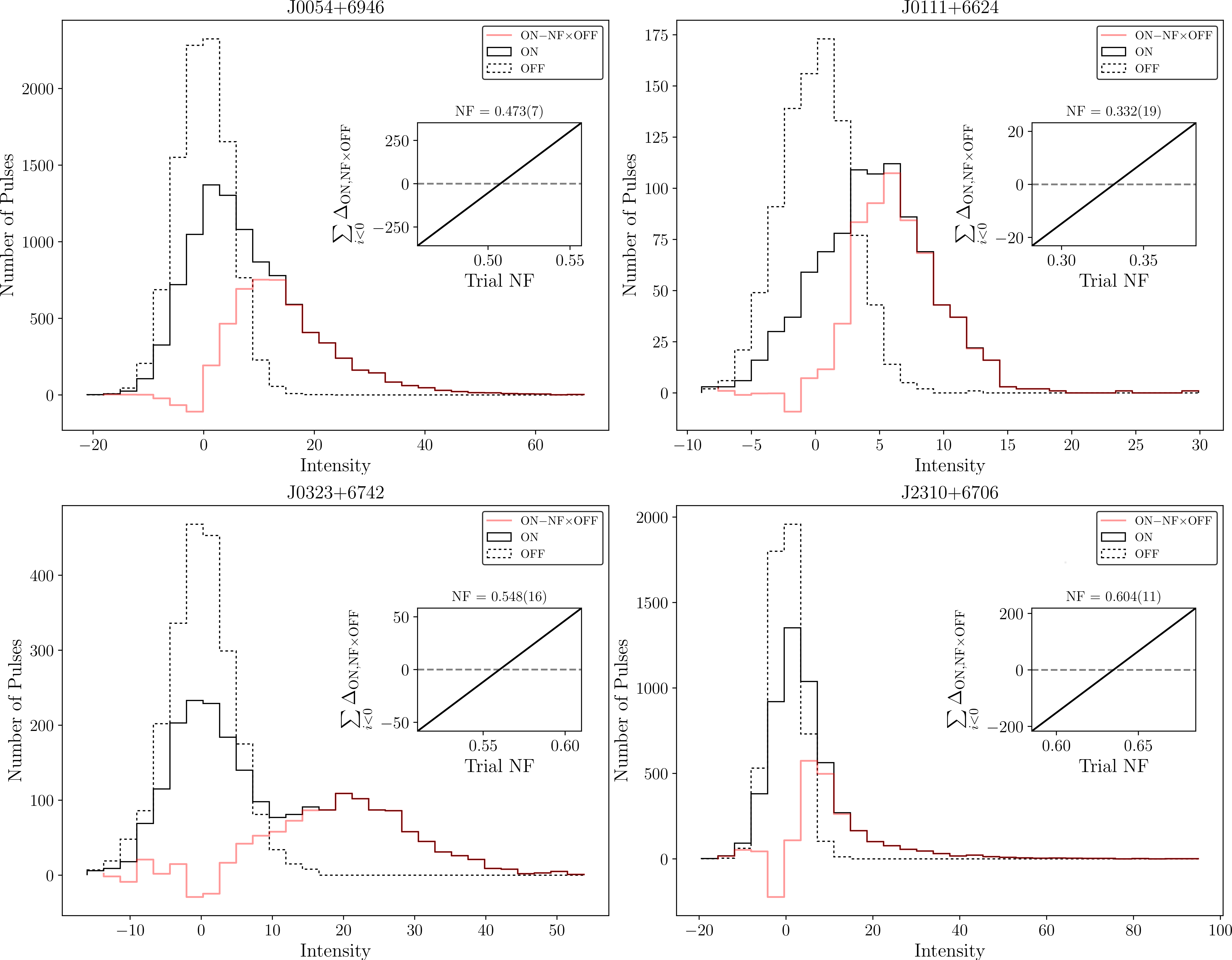}
  \caption{Histograms of summed intensities in ON/OFF windows for PSRs
    J0054+6946 ({\it a}), J0111+6624 ({\it b}), J0323+6742 ({\it c}),
    and J2310+6706 ({\it d}) are shown with solid/dashed lines,
    respectively. Inset in each panel: plots showing the difference,
    ON$-$NF$\times$OFF (summed over bins with intensity $<0$) versus
    trial NF values. The red line shows ON$-$NF$\times$OFF for the best
    NF value (where the summed difference is closest to zero).}
\label{fig:hist1}
\end{figure*}

\begin{figure*}[t!]
  \centering
  \includegraphics[height=0.7\textwidth]{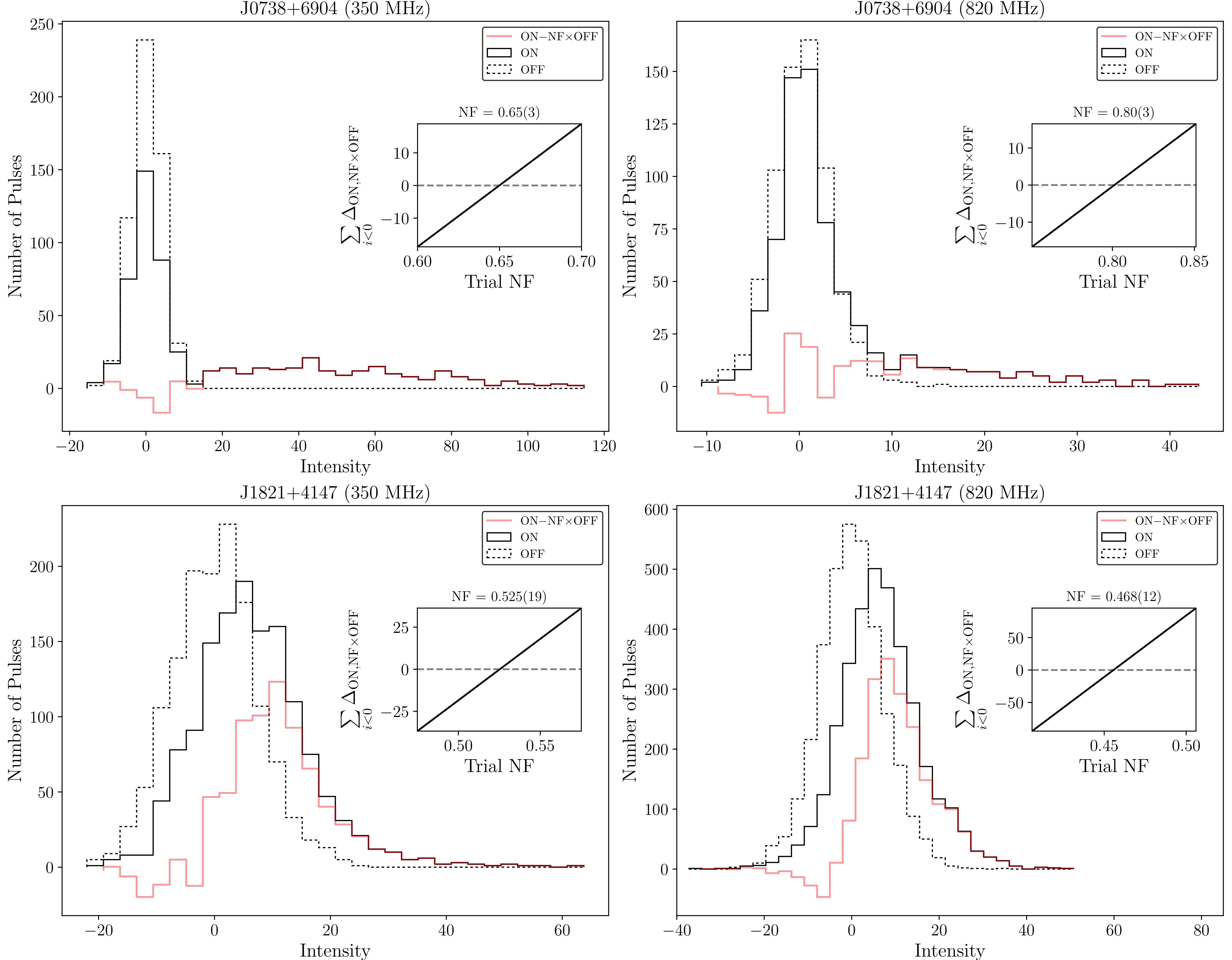}
  \caption{Histograms of summed intensities in ON/OFF windows for PSRs
    J0738+6904 (350~MHz) ({\it a}), J0738+6904 (820~MHz) ({\it b}),
    J1821+4147 (350~MHz) ({\it c}), and J1821+4147 (820~MHz) ({\it d})
    are shown with solid/dashed lines, respectively. Inset in each panel:
    plots showing the difference, ON$-$NF$\times$OFF (summed over bins
    with intensity $<0$) versus trial NF values. The red line shows
    ON$-$NF$\times$OFF for the best NF value (where the summed
    difference is closest to zero).}
  \label{fig:hist2}
\end{figure*}

To conduct an initial census of nulling pulsars among the discoveries
described in this paper, we visually inspected each pulsar's discovery
plot and took note of sources that exhibited obvious intensity
variations as a function of time, resulting in 19 nulling
candidates. We followed a procedure similar to that described in
\citet{rit76} to investigate which (if any) of these pulsars exhibited
measurable nulling behavior. Here, we describe the procedure used to
process and analyze data for each candidate in order to identify
nulling pulsars in this sample.

Once this initial selection was made, we used timing data (at both
$350$ and $820\; \MHz$) on all of the candidate nullers to confirm or
reject them as nulling pulsars and measure nulling fractions where
appropriate.  Subbanded timing data were dedispersed and folded modulo
the pulsar's spin period, resulting in files containing
sub-integrations equivalent in length to the pulsar's spin period and
64 frequency channels across the band. RFI was removed interactively
with {\tt pazi}, part of the {\sc PSRCHIVE} pulsar processing software
package\footnote{\url{http://psrchive.sourceforge.net/}}
\citep{hvm04}. After downsampling all cleaned detections in time and
frequency, the resulting folded profile was used to determine
ON/OFF-pulse windows of equal size. In most cases, ON/OFF windows
spanned $\lesssim15$~bins, or about 5\% of a full rotation (256
bins). The ON window was centered on the ON-pulse region and the OFF
window was fixed 100 bins away, sampling baseline noise.

Cleaned data were downsampled in frequency and remaining bins outside
both ON/OFF windows were used to subtract a DC offset (mean value)
from single pulses; any remaining low-frequency noise was also removed
during this stage by fitting out a 6th-order polynomial from the
baseline. Examples of resulting single pulse intensities (and folded
profiles) are shown in Figure \ref{fig:singlepulses}. Summed
intensities were recorded for each single pulse in both ON and OFF
windows, then binned over intensity to create histograms (see Figures
\ref{fig:hist1} and \ref{fig:hist2}) for each window. In order to
identify pulsars exhibiting some form of nulling behavior, we looked
for sources with ON histograms that showed both measurable single
pulse emission (a tail or positive distribution of ON intensities) and
a distribution of null pulses centered on zero intensity, inside the
envelope of the OFF histogram (see panel for PSR J0323+6742 in Figure
\ref{fig:hist1} for a canonical example).

Of the 19 candidates for which we carried out this analysis, eight
pulsars (group A; PSRs J0612+3721, J1859+7654, J1941+4320, J1954+4357,
J2137+6428, J2228+6447, J2312+6931, and J2351+8533) were too weak to
see single pulses and, therefore, results were inconclusive. Five
others (group B; PSRs J0137+6349, J0335+6623, J0944+4106, J1059+6459,
and J1647+6609) showed obvious single pulse emission, but no null
distribution was apparent in any of their ON histograms using existing
data. Based on our analysis, pulsars from group A have single pulse
emission below our detection threshold, but because of intensity
variability apparent in their discovery plots, we may be able to carry
out a similar analysis by summing groups of single pulses \citep[as
  in][]{wmj07} and looking for longer nulls. Pulsars from group B may
have extremely low NFs, which may be detectable with extended
datasets.

The remaining six pulsars from the original candidate list (PSRs
J0054+6946, J0111+6624, J0323+6742, J0738+6904, J1821+4147, and
J2310+6706) were identified as new nulling pulsars (see Figure
\ref{fig:singlepulses} for profiles and examples of their single pulse
behavior. Using their ON/OFF histograms (see Figures \ref{fig:hist1}
and \ref{fig:hist2}), we computed preliminary NFs and found values
between 0.33--0.80. For PSRs J0738+6904 and J1821+4147, timing data
from multiple frequencies ($350$ and $820\; \MHz$) allowed us to
compare nulling behavior across frequency; based on preliminary
results shown in Figure \ref{fig:hist2}, we found similar ON
distributions and NF values, independent of observing frequency, as
also found by \citet{gjk+14} on 3 nulling pulsars.

An analysis that will select nulling pulsars using all existing timing
data with an improved method that uses Gaussian mixture models and
accounts for the effects of scintillation is underway.  Also of
interest in a future study would be a detailed comparison between
these new sources and the rest of the known nulling population as well
as predictions about the underlying nulling population, given the
relatively complete and unbiased nature of the GBNCC survey.

\section{Optical Constraints}
\label{sec:optical}

\begin{deluxetable*}{lcccccc}[t]
 \tablewidth{0pt}
\tablecaption{Optical Counterpart Limits \label{tab:optlims}}
\tablehead{\colhead{Source} & \colhead{Companion Mass$^{\rm a}$} & \colhead{Companion Type} & \colhead{T$_{\rm eff}$} & \colhead{Constraining Filter} & \colhead{WD Age}  \\
 & \colhead{(M$_\odot$)} & & \colhead{(K)} & (g/r/i/z) & \colhead{(Gyr)}}
\startdata
J0740+6620 & \phn0.20 & He WD & \phn$<4200$ & i & \phn$>3.2$ \\
 & \phn0.23 & & \phn$<4400$ & & \phn$>4.0$ & \\
J1125+7819 & \phn0.29 & He WD & \phn$<4400$ & i & \phn$>3.3$ \\
 & \phn0.33 & & \phn$<3500$ & & \phn$>5.6$  & \\
J1938+6604 & \phn0.87 & CO WD & \phn$<20000$ & g & \phn$>0.179$ \\
 & \phn1.04 & & \phn$<25000$ & & \phn$>0.118$ & \\
\enddata
\tablenotetext{a}{Minimum and median companion masses displayed for
  each source, assuming $m_{\rm p}=1.35$\,\Msun and orbital
  inclination angles of 90$^{\circ}$ and 60$^{\circ}$, respectively.}
\tablecomments{We use the larger of the DM-inferred distance from
  either the NE2001 or YMW16 models in setting limits.}
\end{deluxetable*}

Each of the MSPs with a binary companion and nearly circular orbit
(see Table \ref{table:binary}) most likely has a WD companion. For all
of them, we looked for optical counterparts using data from the
PanSTARRS $3\pi$ Steradian Survey \citep{cmm+16}. By manually
inspecting individual ({\tt grizy}) bands from the PanSTARRS survey's
PS1 data release, we found a counterpart for PSR J1641+8049
(\S\ref{sec:J1641_optical}), but none were found for the three other
sources. For non-detections, we used the minimum/median companion
masses derived from the systems' binary parameters (see Table
\ref{tab:optlims}) and the average 5-$\sigma$ magnitude lower limits
for the PS1 {\tt grizy} bands \citep[23.3, 23.2, 23.1, 22.3, and 21.4,
  respectively; ][]{cmm+16} to place constraints on the properties of
each MSP's WD companion. Based on computed minimum companion masses,
PSRs J0740+6620 and J1125+7819 likely have He WD companions and PSR
J1938+6604, a CO WD companion.

Reddening was estimated using a 3-D map of interstellar dust reddening
by \citet{gsf+15}, the pulsar's sky position and $D_{\rm DM}$ (see
Tables 4 and 5; we used the larger of the DM-inferred distances from
the NE2001 or YMW16 models to calculate the most conservative
limits). These values were converted to extinctions in each PS1 band
using Table 6 from \cite{sf11}. Resulting magnitude limits were
translated to upper limits on WD effective temperatures ($T_{\rm
  eff}$) using models for WD mass, radius and temperature from
\cite{imt+16}. The best WD $T_{\rm eff}$ constraints are listed for
each pulsar's minimum/median companion mass in Table
\ref{tab:optlims}, along with the PS1 bands providing those
constraints, and modeled WD cooling ages \citep{imt+16}. Because of
its significantly higher companion mass, we used different
models\footnote{\url{http://www.astro.umontreal.ca/$\sim$bergeron/CoolingModels/}}
\citep{tbg11,bwd+11} to place constraints on PSR J1938+6604's CO WD,
but otherwise followed the same procedure described here.

Based on mass and temperature constraints, both PSRs J0740+6620 and
J1125+7819 have cool He WD companions; for these systems, a
correlation between $P_{\rm b}$ and WD mass is expected
\citep{sav87,ts99} and the $(P_{\rm b},m_{\rm WD})$-relationship
predicts companion masses of 0.25\,\Msun and 0.28\,\Msun,
respectively.  These are consistent with the minimum companion masses
for these systems.

\section{Conclusion}
\label{sec:conc}

We present here complete timing solutions for \npsrs\ new pulsars
discovered by the GBNCC survey.  The highlights include two new PTA
pulsars that are being used in an effort to detect low-frequency GWs,
several intermittent pulsars, a mode changing pulsar, and five binary
pulsars.  Among the binary pulsars are a new DNS system, IMBP, and
black widow.

These results demonstrate the importance of long-term pulsar timing,
as many properties, such as spin-down and proper motion, can only be
measured with data spanning a year or more.  We continue to observe
select sources to improve the measurements of proper motion and some
binary parameters.  We also are observing approximately 100 additional
pulsars discovered in the GBNCC survey, which will be presented in
future work.  Given our current estimate of survey yield, we expect to
discover an additional few dozen long-period pulsars and of order
5--10 MSPs before the survey is completed.

\acknowledgements

\emph{Acknowledgements:} We would like to thank an anonymous referee
whose comments improved the quality of this work.  The Green Bank
Observatory is a facility of the National Science Foundation (NSF)
operated under cooperative agreement by Associated Universities, Inc.
This paper is based in part on data obtained with the International
LOFAR Telescope (ILT). LOFAR \citep{vwg+13} is the Low Frequency Array
designed and constructed by ASTRON. It has facilities in several
countries, that are owned by various parties (each with their own
funding sources), and that are collectively operated by the ILT
foundation under a joint scientific policy.  These results also made
use of Lowell Observatory's Discovery Channel Telescope. Lowell
operates the DCT in partnership with Boston University, Northern
Arizona University, the University of Maryland, and the University of
Toledo. Partial support of the DCT was provided by Discovery
Communications. LMI was built by Lowell Observatory using funds from
the NSF (AST-1005313).  This paper includes data taken at The McDonald
Observatory of The University of Texas at Austin.  IRAF is distributed
by the National Optical Astronomy Observatory, which is operated by
the Association of Universities for Research in Astronomy (AURA) under
cooperative agreement with the NSF. J.K.S., M.E.D., F.A.J., D.L.K.,
M.A.M., S.M.R., K.S., and X.S.\ are supported by the NANOGrav NSF
Physics Frontiers Center award number 1430284.  M.E.D.\ also
acknowledges support from NSF award number AST-1312822.  V.I.K.\ and
J.W.T.H.\ acknowledge funding from an NWO Vidi fellowship and the
European Research Council under the European Union's Seventh Framework
Programme (FP/2007-2013) / ERC Starting Grant agreement nr. 337062
("DRAGNET").  V.M.K.\ receives support from an NSERC Discovery Grant
and Accelerator Supplement, from the Centre de Recherche en
Astrophysique du Qu\'{e}bec, an R.\ Howard Webster Foundation
Fellowship from the Canadian Institute for Advanced Research, the
Canada Research Chairs Program, and the Lorne Trottier Chair in
Astrophysics and Cosmology.  M.A.M.\ and B.C.\ were supported by NSF
awards AST-1327526 and OIA-1458952.  S.M.R.\ is a Canadian Institute
for Advanced Research Senior Fellow.  I.H.S.\ and pulsar research at
UBC are supported by an NSERC Discovery Grant and by the Canadian
Institute for Advanced Research.  J.vL.\ receives funding from the
European Research Council under the European Union's Seventh Framework
Programme (FP/2007-2013)/ERC Grant Agreement n.\ 617199.

\facility{GBT (GUPPI), LOFAR, \textit{Fermi} (LAT), DCT (LMI),
  McDonald Observatory 1-m (Sinistro)}

\software{\texttt{PRESTO}
  (\url{https://github.com/scottransom/presto}), \texttt{PSRCHIVE}
  \citep{hvm04}, \texttt{TEMPO} (\url{http://tempo.sourceforge.net/}),
  \texttt{PINT} (\url{https://github.com/nanograv/PINT}),
  \texttt{IRAF} \citep{tod86,tod93}, \texttt{SCAMP}
  \citep{ba96}}

\figsetstart
\figsetnum{8}
\figsettitle{Timing Residuals for \npsrs\ GBNCC Pulsars}

\figsetgrpstart
\figsetgrpnum{8.1}
\figsetgrptitle{Timing Residuals for GBNCC Pulsars (Page 1)}
\figsetplot{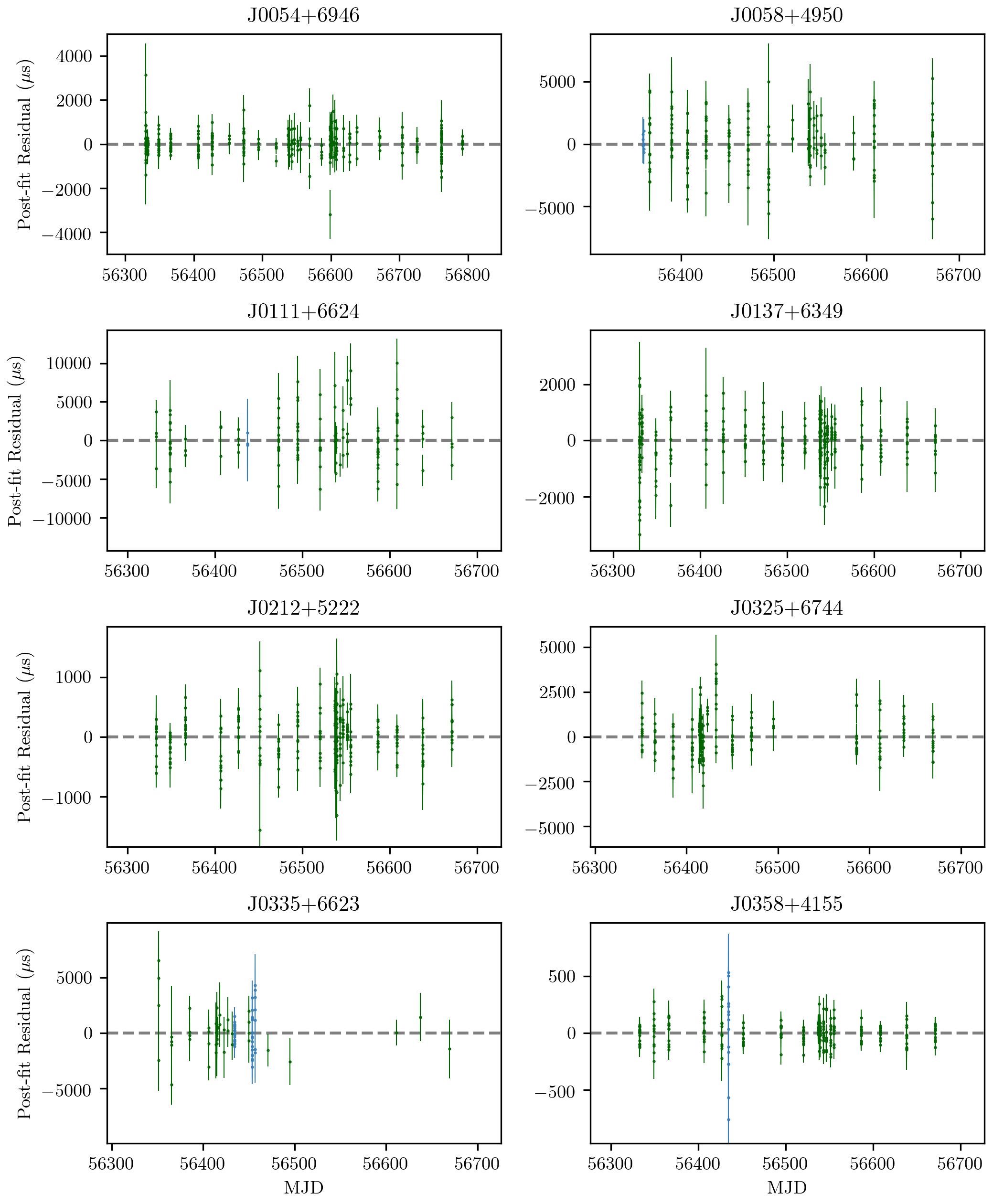}
\figsetgrpnote{Post-fit timing residuals for the \npsrs\ pulsars
  presented here.  Colors correspond to different observing bands:
  $148\; \MHz$ (blue) $350\; \MHz$ (orange), $820\; \MHz$ (dark
  green), $1500\; \MHz$ (pink), and $2000\; \MHz$ (brown).  Note that
  vertical and horizontal scales differ for each pulsars.}
\figsetgrpend

\figsetgrpstart
\figsetgrpnum{8.2}
\figsetgrptitle{Timing Residuals for GBNCC Pulsars (Page 2)}
\figsetplot{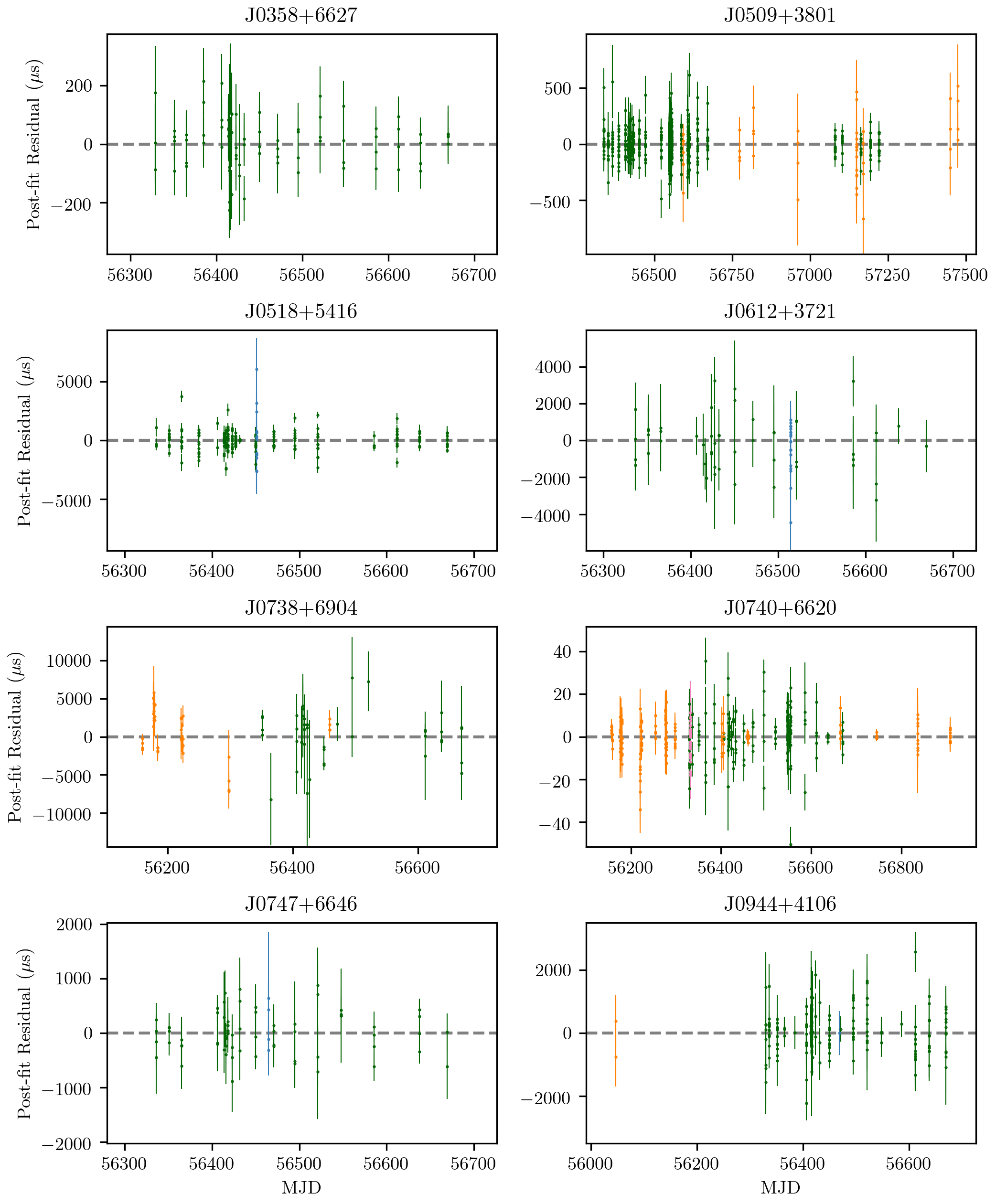}
\figsetgrpnote{Post-fit timing residuals for the \npsrs\ pulsars
  presented here.  Colors correspond to different observing bands:
  $148\; \MHz$ (blue) $350\; \MHz$ (orange), $820\; \MHz$ (dark
  green), $1500\; \MHz$ (pink), and $2000\; \MHz$ (brown).  Note that
  vertical and horizontal scales differ for each pulsars.}
\figsetgrpend

\figsetgrpstart
\figsetgrpnum{8.3}
\figsetgrptitle{Timing Residuals for GBNCC Pulsars (Page 3)}
\figsetplot{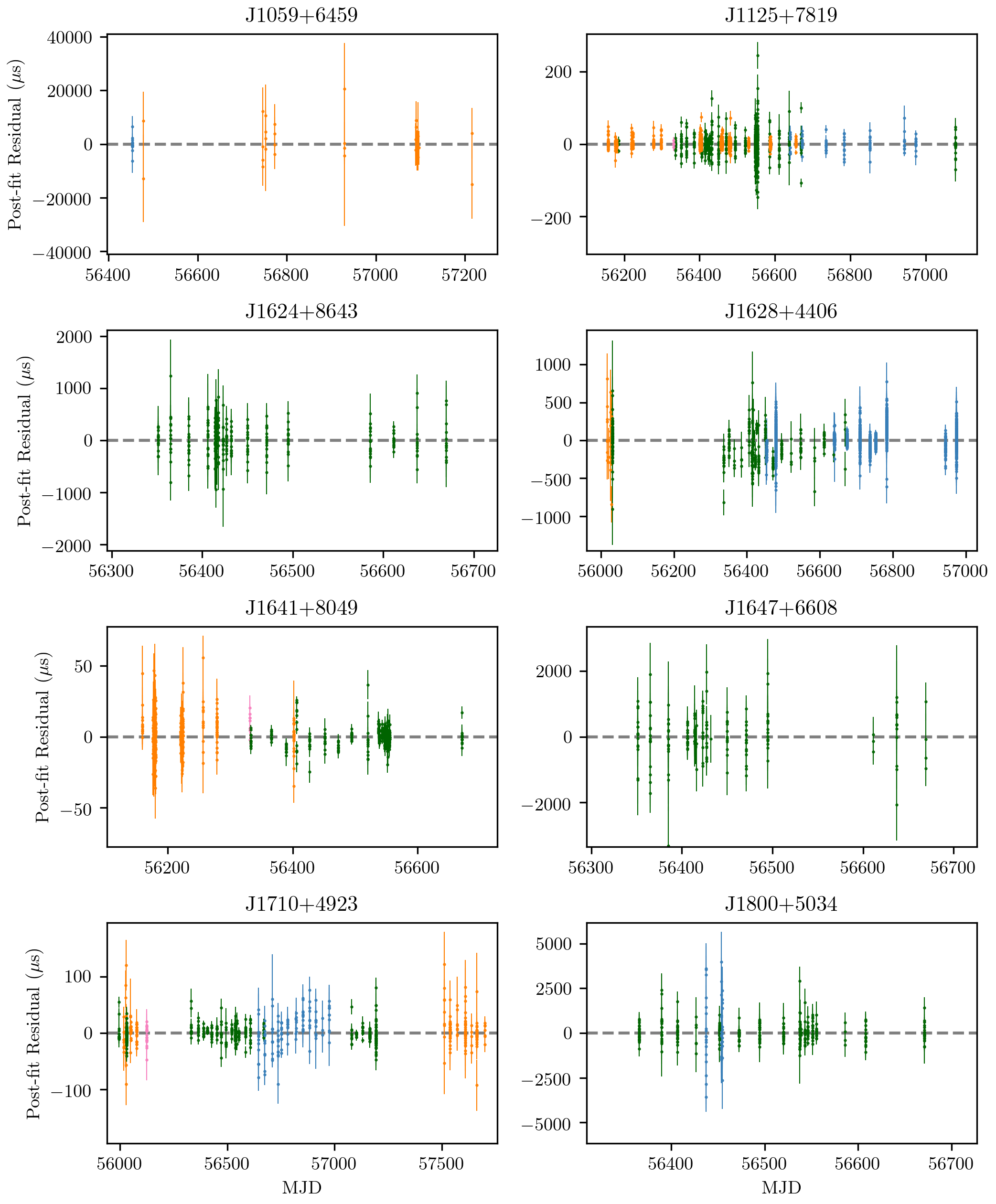}
\figsetgrpnote{Post-fit timing residuals for the \npsrs\ pulsars
  presented here.  Colors correspond to different observing bands:
  $148\; \MHz$ (blue) $350\; \MHz$ (orange), $820\; \MHz$ (dark
  green), $1500\; \MHz$ (pink), and $2000\; \MHz$ (brown).  Note that
  vertical and horizontal scales differ for each pulsars.}
\figsetgrpend

\figsetgrpstart
\figsetgrpnum{8.4}
\figsetgrptitle{Timing Residuals for GBNCC Pulsars (Page 4)}
\figsetplot{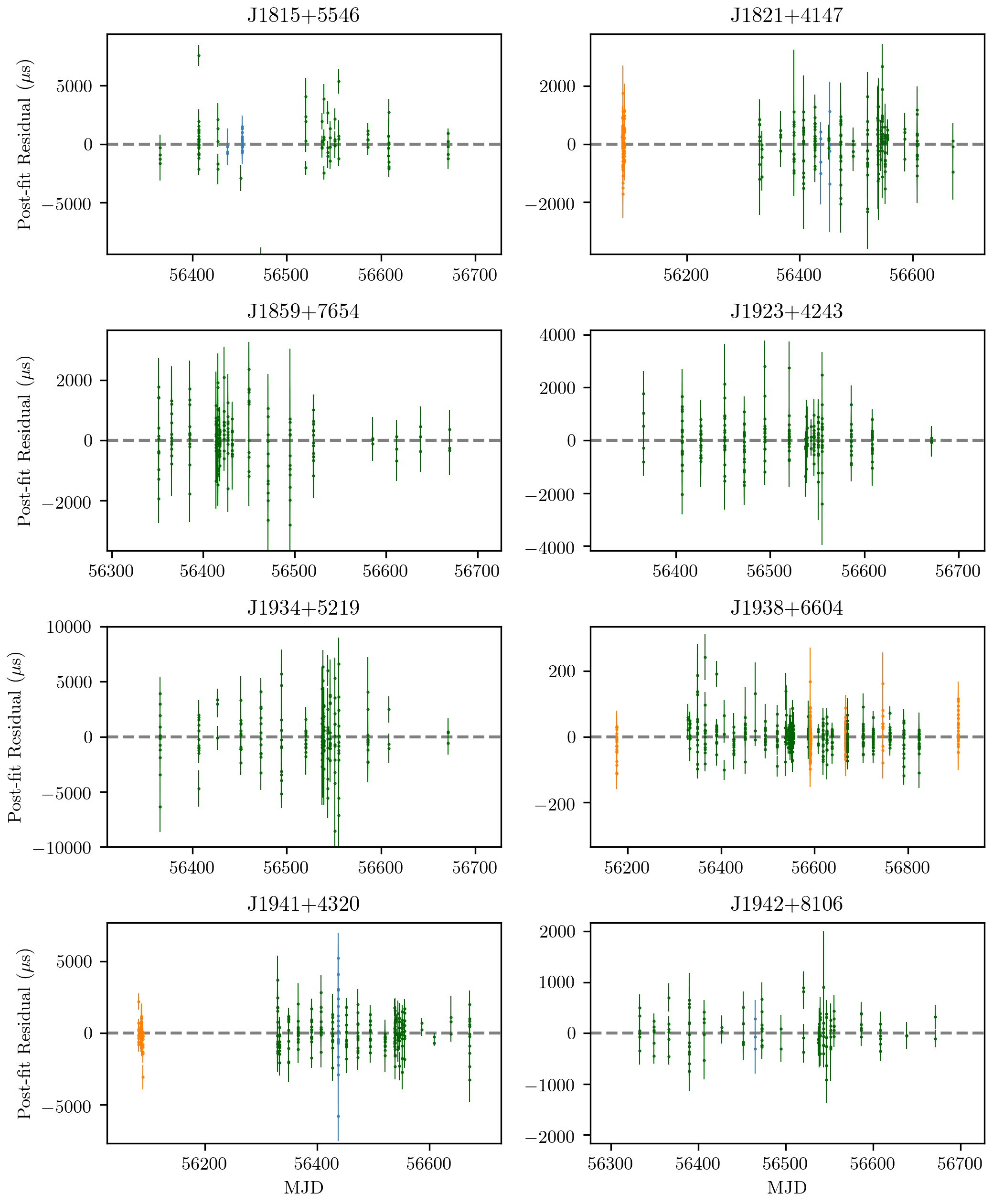}
\figsetgrpnote{Post-fit timing residuals for the \npsrs\ pulsars
  presented here.  Colors correspond to different observing bands:
  $148\; \MHz$ (blue) $350\; \MHz$ (orange), $820\; \MHz$ (dark
  green), $1500\; \MHz$ (pink), and $2000\; \MHz$ (brown).  Note that
  vertical and horizontal scales differ for each pulsars.}
\figsetgrpend

\figsetgrpstart
\figsetgrpnum{8.5}
\figsetgrptitle{Timing Residuals for GBNCC Pulsars (Page 5)}
\figsetplot{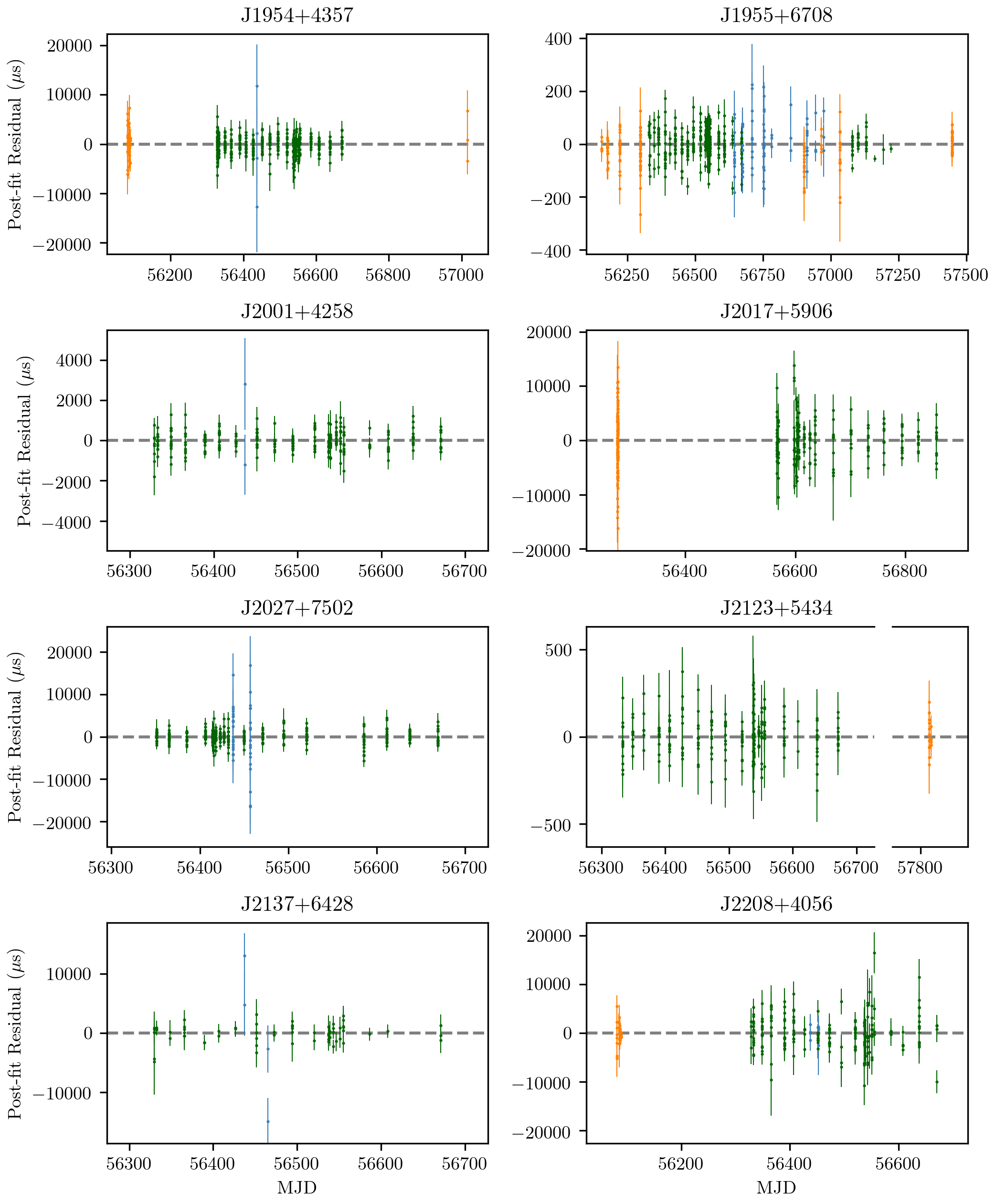}
\figsetgrpnote{Post-fit timing residuals for the \npsrs\ pulsars
  presented here.  Colors correspond to different observing bands:
  $148\; \MHz$ (blue) $350\; \MHz$ (orange), $820\; \MHz$ (dark
  green), $1500\; \MHz$ (pink), and $2000\; \MHz$ (brown).  Note that
  vertical and horizontal scales differ for each pulsars.}
\figsetgrpend

\figsetgrpstart
\figsetgrpnum{8.7}
\figsetgrptitle{Timing Residuals for GBNCC Pulsars (Page 7)}
\figsetplot{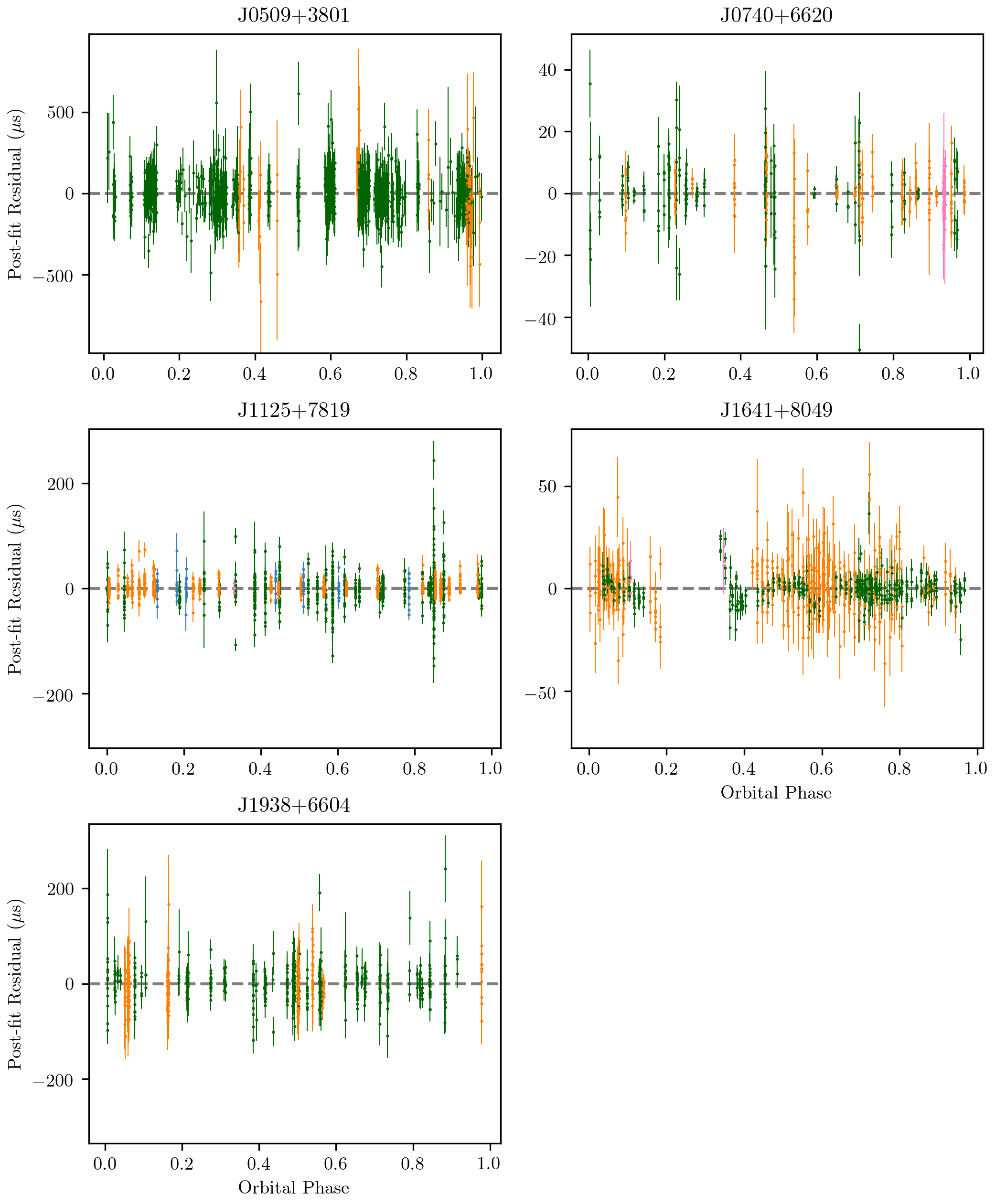}

\figsetgrpnote{Post-fit timing residuals as a function of orbital
  phase for the binary pulsars presented here.  Colors correspond to
  different observing bands: $148\; \MHz$ (blue) $350\; \MHz$
  (orange), $820\; \MHz$ (dark green), $1500\; \MHz$ (pink), and
  $2000\; \MHz$ (brown).  Note that vertical scale differs for each
  pulsars.}

\figsetgrpend

\figsetend

\begin{comment}
\begin{figure*}[p!]
  \centering
  \includegraphics[width=0.9\textwidth]{residuals_page1.png}
    \caption{Post-fit timing residuals for the \npsrs\ pulsars
      presented here.  Colors correspond to different observing bands:
      $148\; \MHz$ (blue) $350\; \MHz$ (orange), $820\; \MHz$ (dark
      green), $1500\; \MHz$ (pink), and $2000\; \MHz$ (brown).  Note
      that vertical and horizontal scales differ for each pulsars. The
      complete figure set (seven images, including timing residuals as
      a function of orbital phase for binary pulsars) is available in
      the online journal.  \label{fig:residuals}}
\end{figure*}
\end{comment}

%\begin{comment}

\begin{figure*}[p!]
  \centering
  \includegraphics[width=0.9\textwidth]{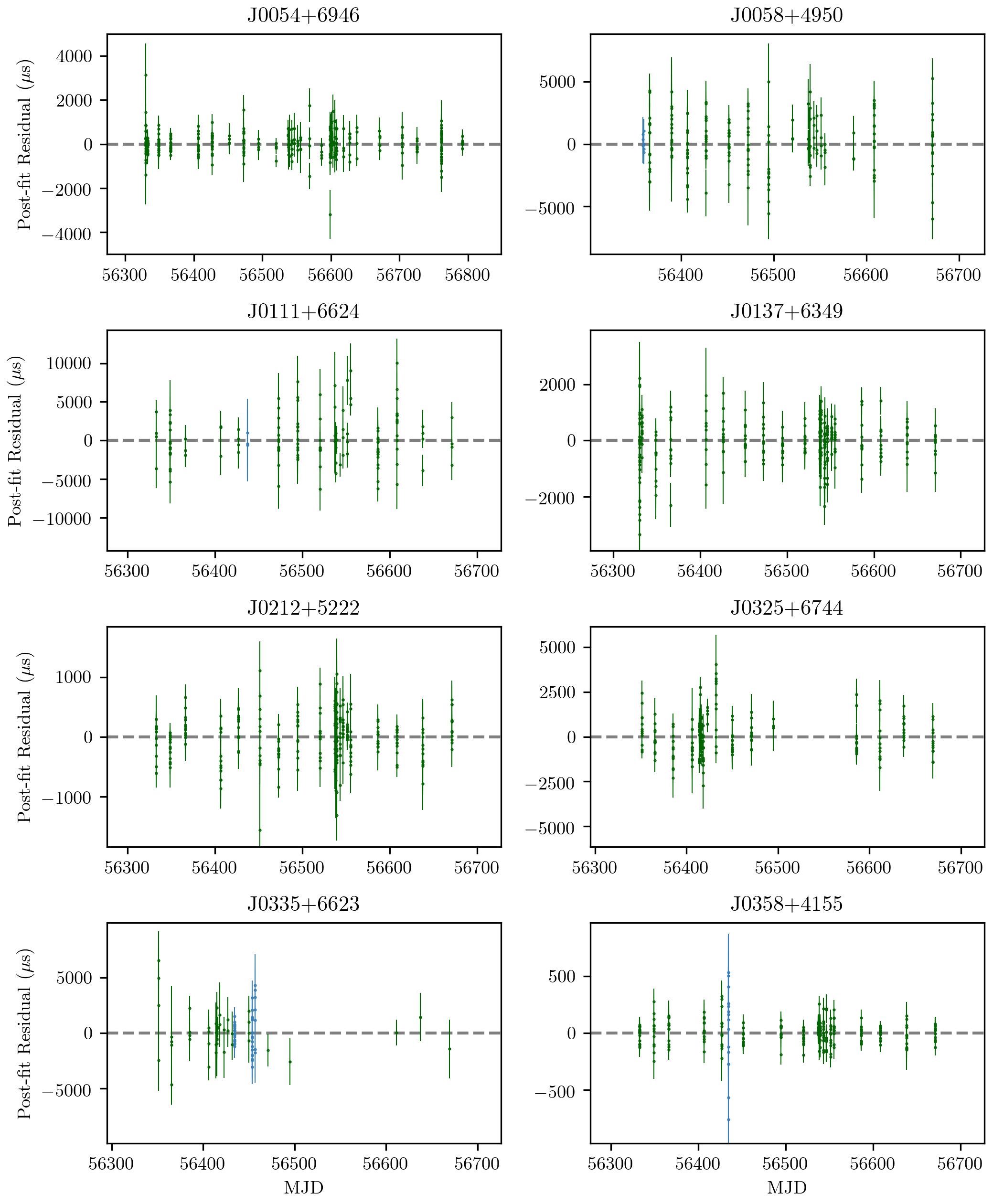}
    \caption{Post-fit timing residuals for the \npsrs\ pulsars
      presented here.  Colors correspond to different observing bands:
      $148\; \MHz$ (blue) $350\; \MHz$ (orange), $820\; \MHz$ (dark
      green), $1500\; \MHz$ (pink), and $2000\; \MHz$ (brown).  Note
      that vertical and horizontal scales differ for each
      pulsars. \label{fig:residuals}}
\end{figure*}
\begin{figure*}[p!]
  \ContinuedFloat
  \centering
  \includegraphics[width=0.9\textwidth]{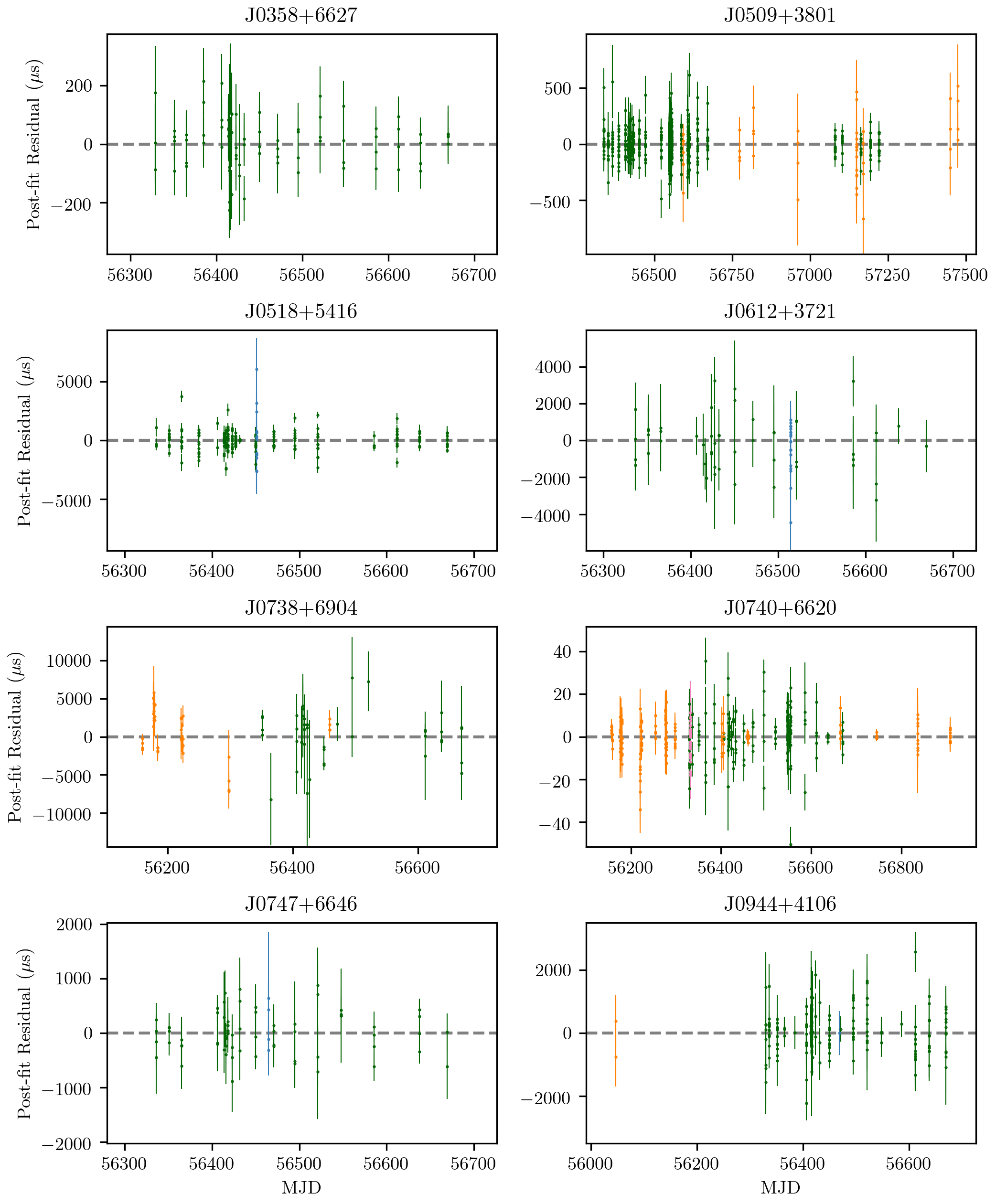}
  \caption{Post-fit timing residuals for the \npsrs\ pulsars presented
    here.  Colors correspond to different observing bands: $148\;
    \MHz$ (blue) $350\; \MHz$ (orange), $820\; \MHz$ (dark green),
    $1500\; \MHz$ (pink), and $2000\; \MHz$ (brown).  Note that
    vertical and horizontal scales differ for each pulsars.}
\end{figure*}
\begin{figure*}[p!]
  \ContinuedFloat
  \centering
  \includegraphics[width=0.9\textwidth]{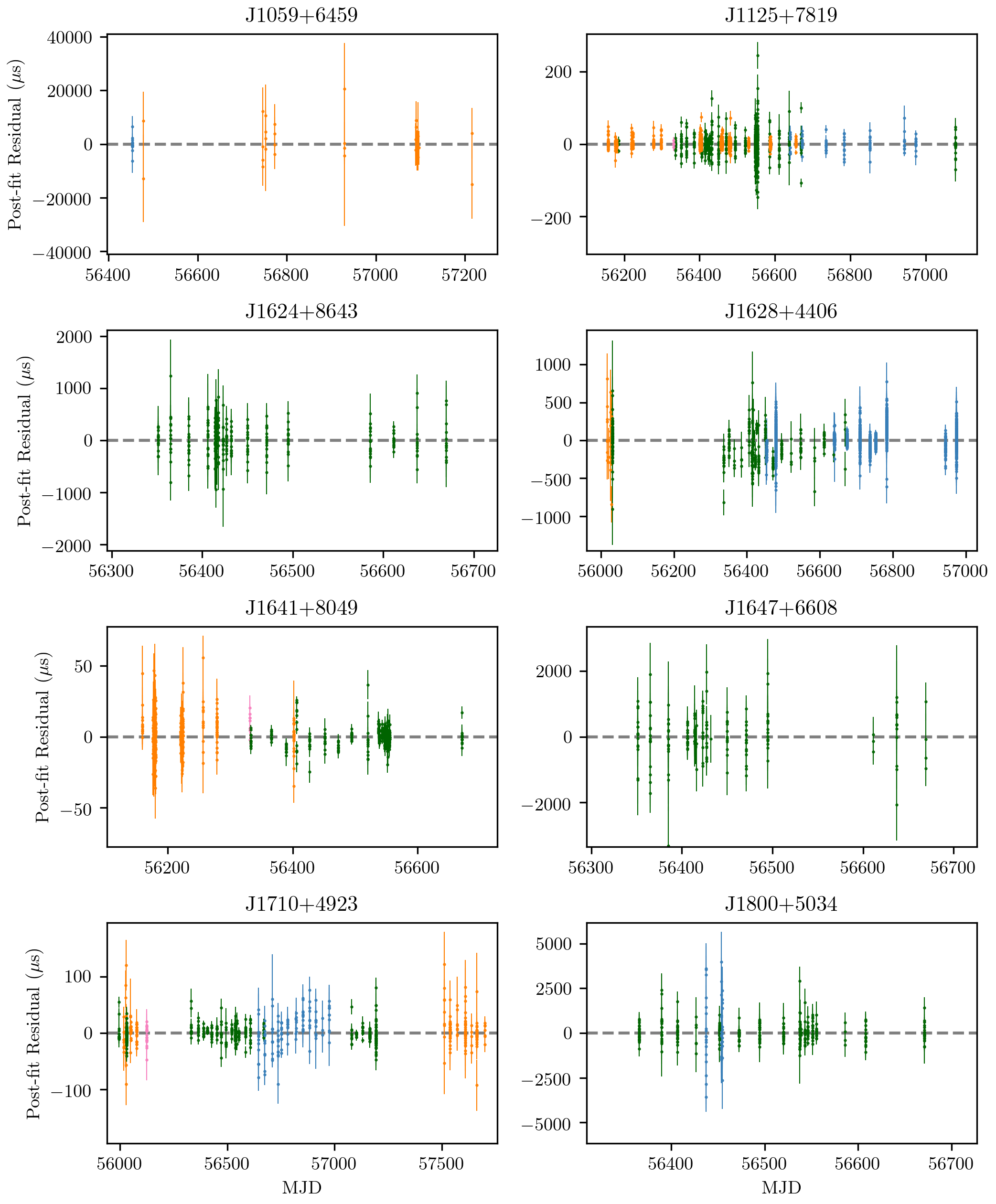}
  \caption{Post-fit timing residuals for the \npsrs\ pulsars presented
    here.  Colors correspond to different observing bands: $148\;
    \MHz$ (blue) $350\; \MHz$ (orange), $820\; \MHz$ (dark green),
    $1500\; \MHz$ (pink), and $2000\; \MHz$ (brown).  Note that
    vertical and horizontal scales differ for each pulsars.}
\end{figure*}
\begin{figure*}[p!]
  \ContinuedFloat
  \centering
  \includegraphics[width=0.9\textwidth]{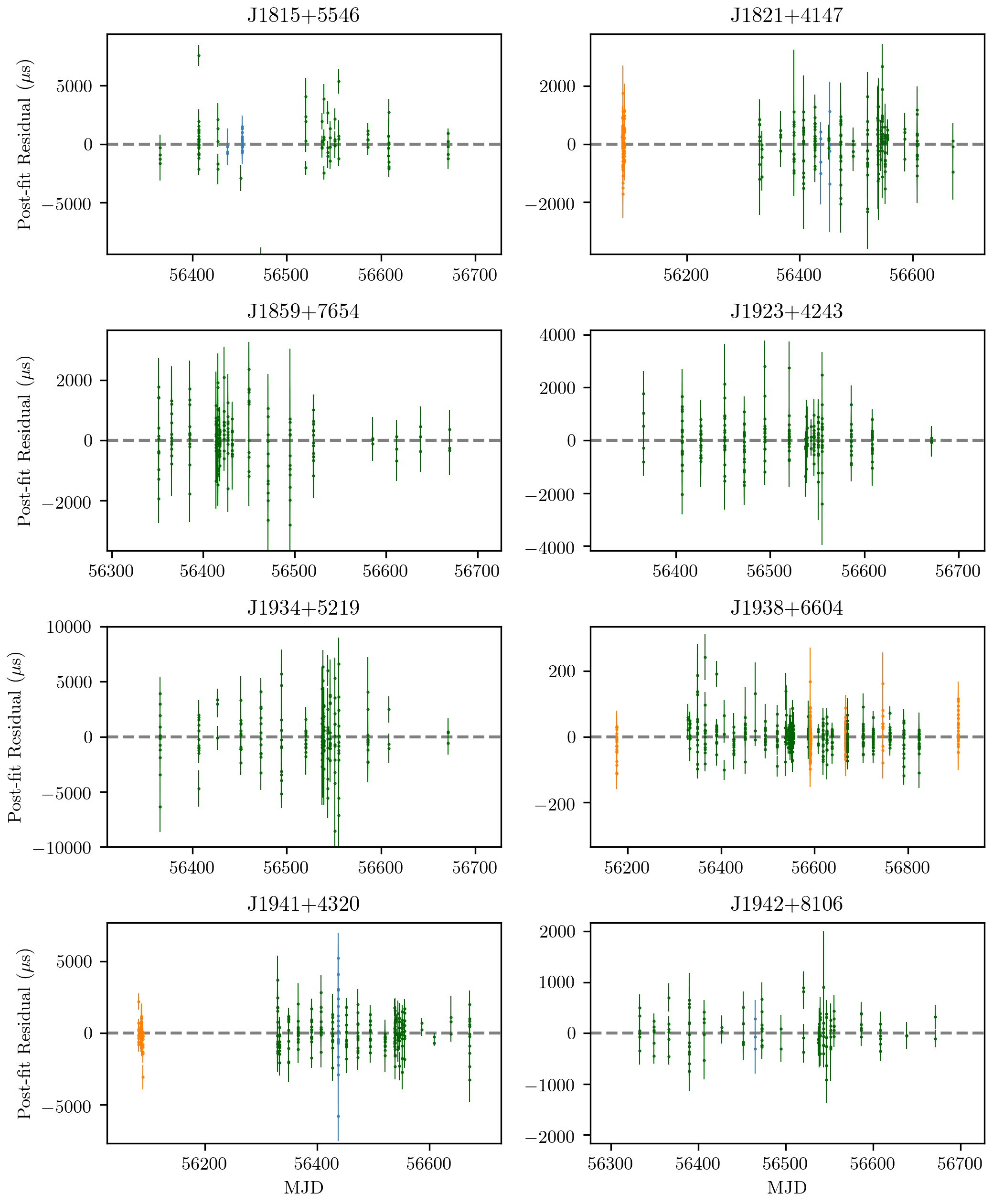}
  \caption{Post-fit timing residuals for the \npsrs\ pulsars presented
    here.  Colors correspond to different observing bands: $148\;
    \MHz$ (blue) $350\; \MHz$ (orange), $820\; \MHz$ (dark green),
    $1500\; \MHz$ (pink), and $2000\; \MHz$ (brown).  Note that
    vertical and horizontal scales differ for each pulsars.}
\end{figure*}
\begin{figure*}[p!]
  \ContinuedFloat
  \centering
  \includegraphics[width=0.9\textwidth]{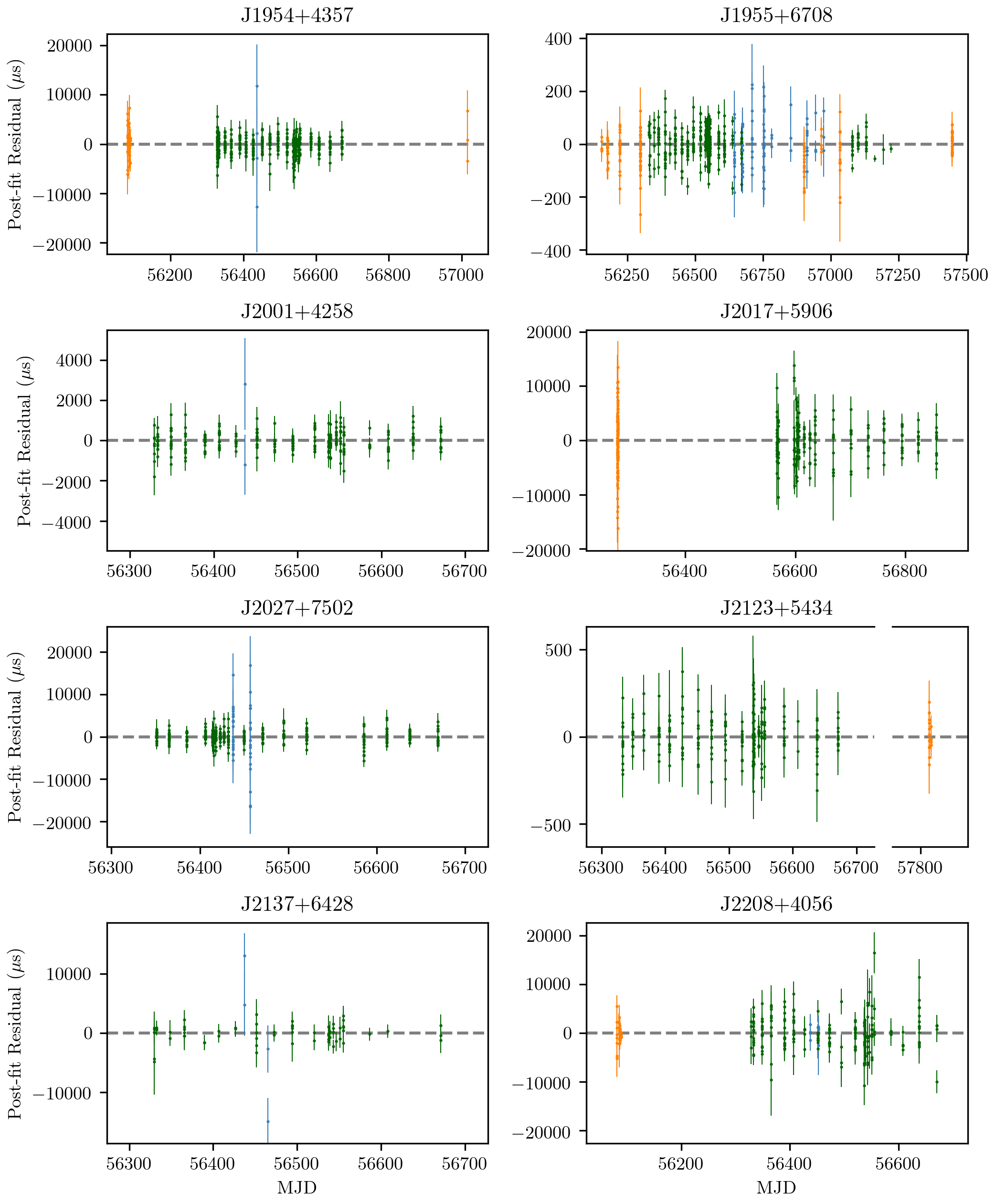}
  \caption{Post-fit timing residuals for the \npsrs\ pulsars presented
    here.  Colors correspond to different observing bands: $148\;
    \MHz$ (blue) $350\; \MHz$ (orange), $820\; \MHz$ (dark green),
    $1500\; \MHz$ (pink), and $2000\; \MHz$ (brown).  Note that
    vertical and horizontal scales differ for each pulsars.}
\end{figure*}
\begin{figure*}[!t]
  \ContinuedFloat
  \centering
  \includegraphics[width=0.9\textwidth]{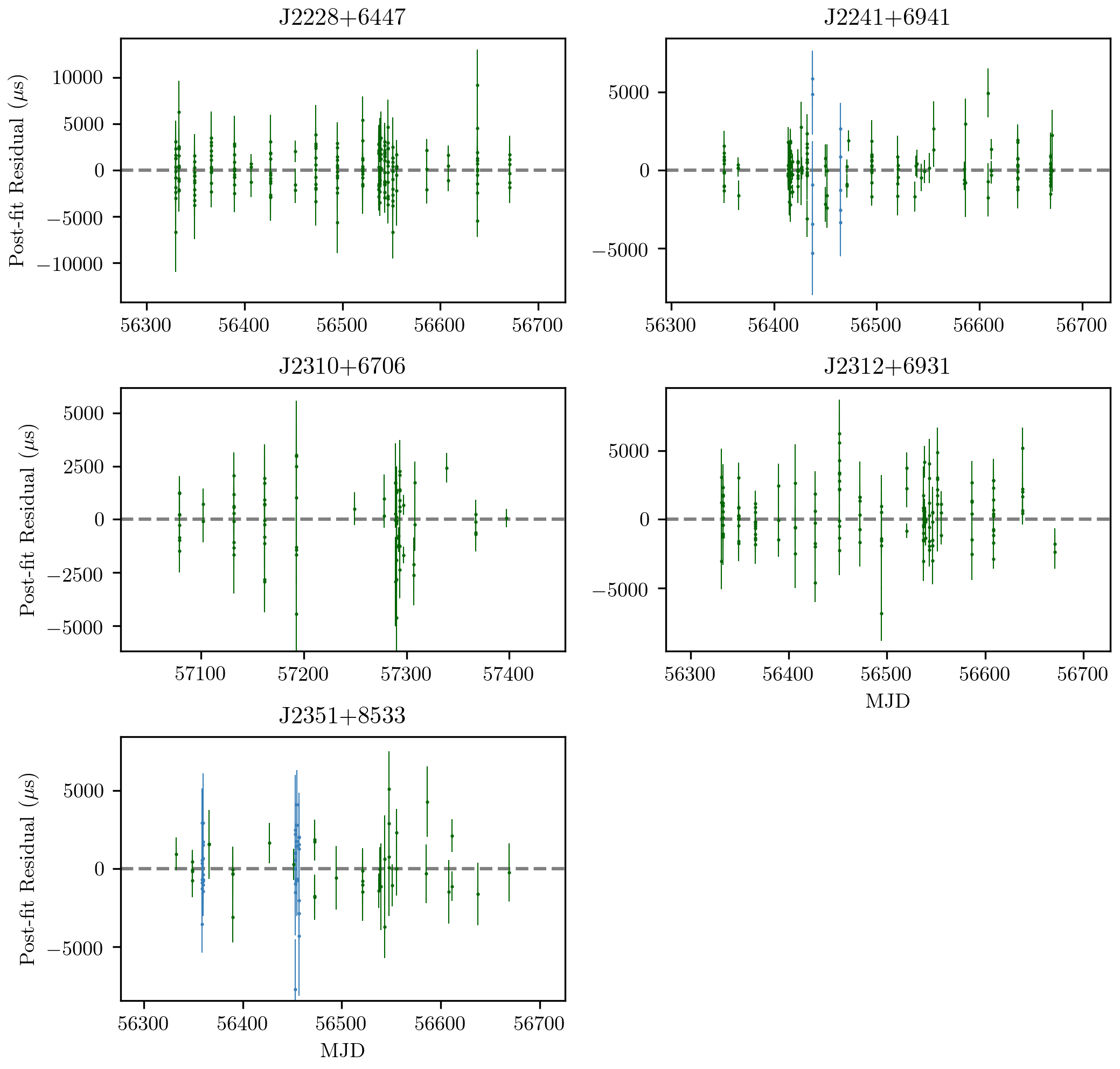}
  \caption{Post-fit timing residuals for the \npsrs\ pulsars presented
    here.  Colors correspond to different observing bands: $148\;
    \MHz$ (blue) $350\; \MHz$ (orange), $820\; \MHz$ (dark green),
    $1500\; \MHz$ (pink), and $2000\; \MHz$ (brown).  Note that
    vertical and horizontal scales differ for each pulsars.}
\end{figure*}

\begin{figure*}[!t]
  \ContinuedFloat
  \centering
  \includegraphics[width=0.9\textwidth]{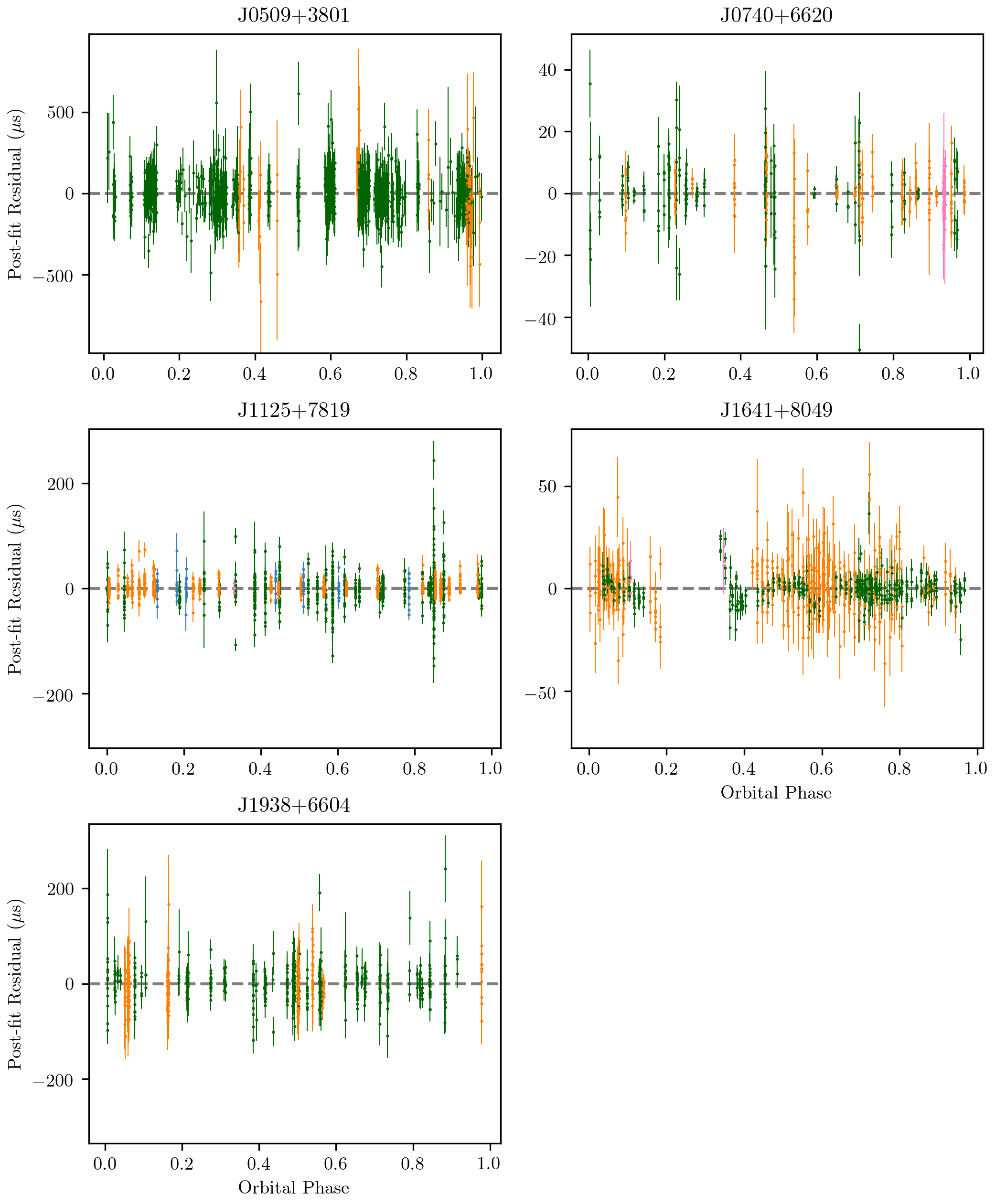}
  \caption{Post-fit timing residuals as a function of orbital phase
    for the binary pulsars presented here.  Colors correspond to
    different observing bands: $148\; \MHz$ (blue) $350\; \MHz$
    (orange), $820\; \MHz$ (dark green), $1500\; \MHz$ (pink), and
    $2000\; \MHz$ (brown).  Note that vertical scale differs for each
    pulsars.}
\end{figure*}

%\end{comment}

\begin{figure*}[p!]
  \centering
  \includegraphics[width=0.9\textwidth]{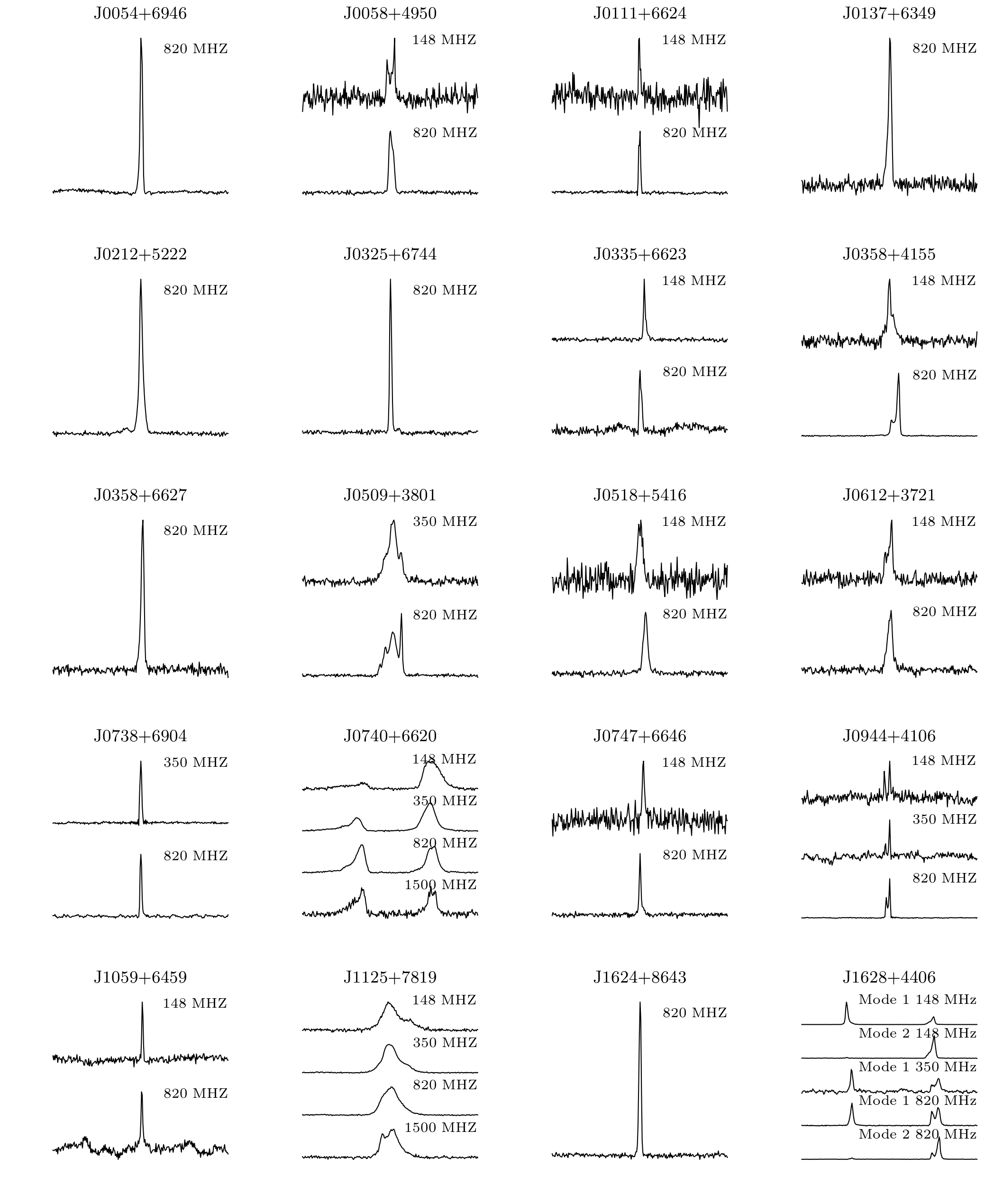}
  \caption{Average pulse profiles for the \npsrs\ pulsars presented here.
    Data from all available observing bands are presented.  All
    profiles have been rotated $180\arcdeg$ from the zero-point pulse
    phase reference.  \label{fig:profiles}}
\end{figure*}
\begin{figure*}[p!]
  \ContinuedFloat
  \centering
  \includegraphics[width=0.9\textwidth]{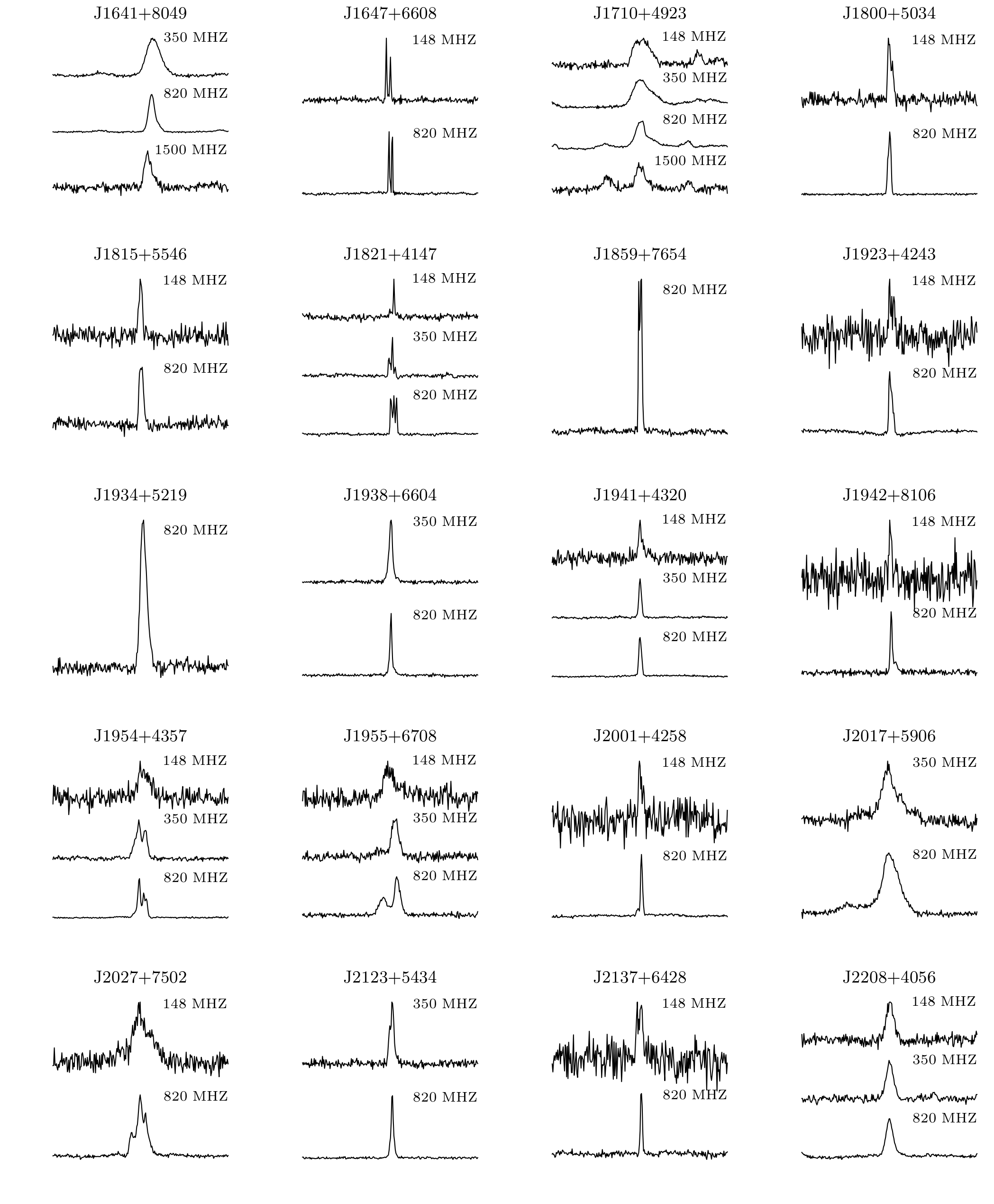}
  \caption{Average pulse profiles for the \npsrs\ pulsars presented here.
    Data from all available observing bands are presented.  All
    profiles have been rotated $180\arcdeg$ from the zero-point pulse
    phase reference.}
\end{figure*}
\begin{figure*}[t!]
  \ContinuedFloat
  \centering
  \includegraphics[width=0.9\textwidth]{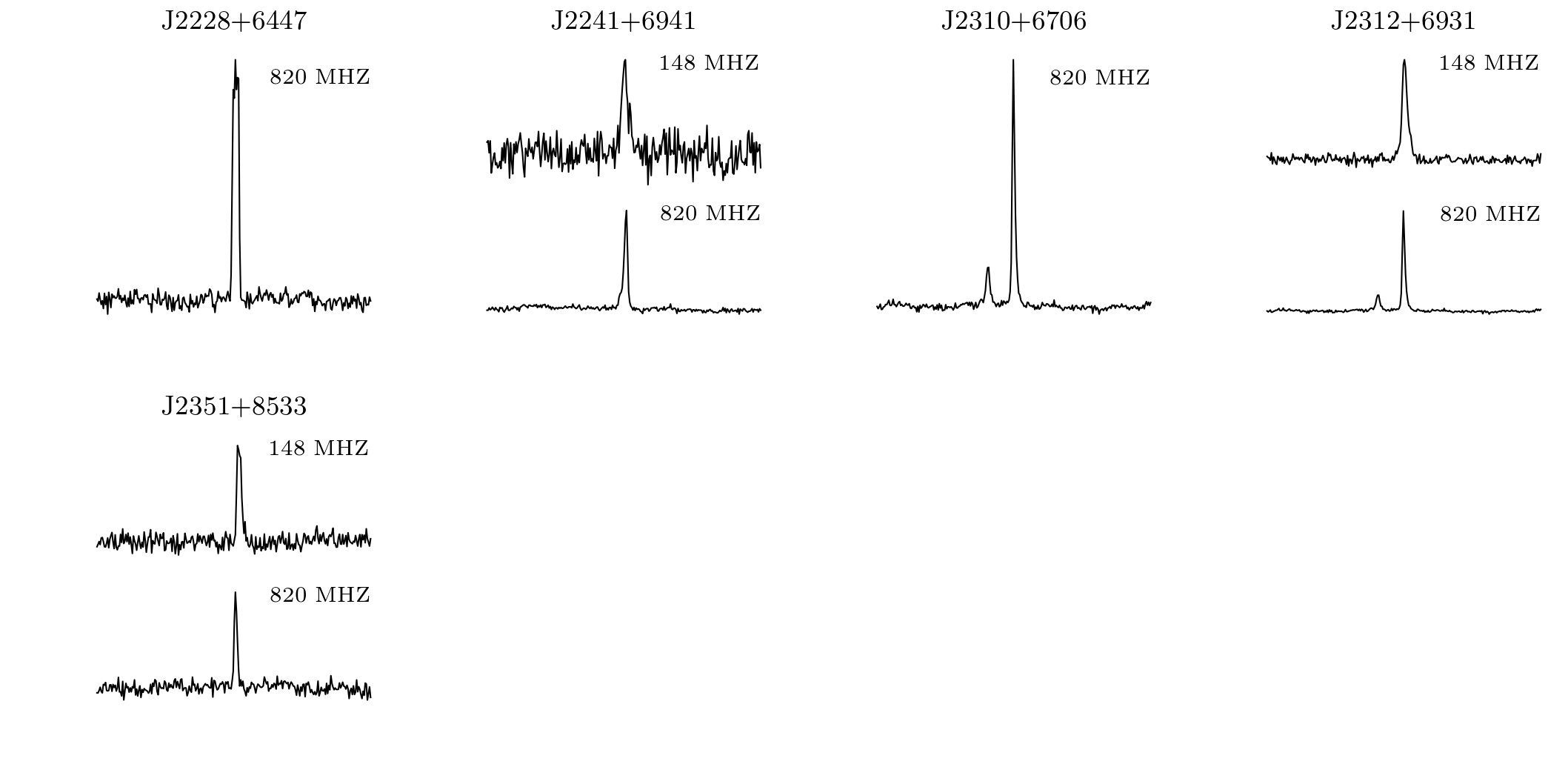}
  \caption{Average pulse profiles for the \npsrs\ pulsars presented here.
    Data from all available observing bands are presented.  All
    profiles have been rotated $180\arcdeg$ from the zero-point pulse
    phase reference.}
\end{figure*}

\clearpage

\bibliographystyle{apj}
\bibliography{references}{}

\end{document}